\documentclass[
reprint,
longbibliography,
superscriptaddress,
preprintnumbers,
nobibnotes,
amsmath,amssymb,
aps,
pra,
floatfix
]{revtex4-1}

\usepackage{gensymb}
\usepackage{mathrsfs}
\usepackage{graphicx}
\usepackage{dcolumn}
\usepackage{bm}
\usepackage{floatrow}
\usepackage{titlesec}

\titleformat*{\section}{\normalsize\bfseries\filcenter}
\titlespacing*{\section}
{0pt}{3.0ex}{2ex}

\usepackage[%
  colorlinks=true,
  urlcolor=blue,
  linkcolor=blue,
  citecolor=blue
]{hyperref}
\usepackage{xcolor}
\usepackage{textcomp}
\usepackage{amsmath}
\usepackage{siunitx}
\usepackage[artemisia]{textgreek}
\usepackage{upgreek}

\usepackage{etoolbox}

\renewcommand{\arraystretch}{1.5}

\newcommand{\unm}{Center for High Technology Materials and Department of Physics and Astronomy, University of New Mexico, Albuquerque, NM, USA}

\begin{document}

\title{Diamond magnetometer enhanced by ferrite flux concentrators}

\author{Ilja Fescenko}
\email{iliafes@gmail.com}
\affiliation{\unm}

\author{Andrey Jarmola}
\affiliation{ODMR Technologies Inc., El Cerrito, CA, USA}
\affiliation{Department of Physics, University of California, Berkeley, CA, USA}

\author{Igor Savukov}
\affiliation{Los Alamos National Laboratory, Los Alamos, NM, USA}

\author{Pauli Kehayias}
\affiliation{\unm}
\affiliation{Sandia National Laboratory, Albuquerque, NM, USA}

\author{Janis Smits}
\affiliation{\unm}
\affiliation{Laser Center of the University of Latvia, Riga, Latvia}

\author{Joshua Damron}
\affiliation{\unm}

\author{Nathaniel Ristoff}
\affiliation{\unm}

\author{Nazanin Mosavian}
\affiliation{\unm}

\author{Victor M. Acosta}
\email{vmacosta@unm.edu}
\affiliation{\unm}


\begin{abstract}
Magnetometers based on nitrogen-vacancy (NV) centers in diamond are promising room-temperature, solid-state sensors. However, their reported sensitivity to magnetic fields at low frequencies (${\lesssim}1~{\rm kHz}$) is presently ${\gtrsim}10~{\rm pT\,s^{1/2}}$, precluding potential applications in medical imaging, geoscience, and navigation. Here we show that high-permeability magnetic flux concentrators, which collect magnetic flux from a larger area and concentrate it into the diamond sensor, can be used to improve the sensitivity of diamond magnetometers. By inserting an NV-doped diamond membrane between two ferrite cones in a bowtie configuration, we realize a $\sim250$-fold increase of the magnetic field amplitude within the diamond. We demonstrate a sensitivity of $\sim0.9~{\rm pT\,s^{1/2}}$ to magnetic fields in the frequency range between $10$ and $1000~{\rm Hz}$, using a dual-resonance modulation technique to suppress the effect of thermal shifts of the NV spin levels. This is accomplished using $200~{\rm mW}$ of laser power and $20~{\rm mW}$ of microwave power. This work introduces a new dimension for diamond quantum sensors by using micro-structured magnetic materials to manipulate magnetic fields.

\end{abstract}

\maketitle

\section{\label{sec:Introduction}Introduction}
Quantum sensors based on nitrogen-vacancy (NV) centers in diamond have emerged as a powerful platform for detecting magnetic fields across a range of length scales~\cite{DEG2017}. At the few-nanometer scale, single NV centers have been used to detect magnetic phenomena in condensed-matter~\cite{CAS2018,ACO2019} and biological~\cite{SCH2014,WU2016} samples. At the scale of a few hundred nanometers, diamond magnetic microscopes have been used to image biomagnetism in various systems, including magnetically-labeled biomolecules~\cite{LOU2019} and cells~\cite{STE2013,GLE2015} and intrinsically-magnetic biocrystals~\cite{FES2019,MCC2019}. At the micrometer scale, diamond magnetometers have detected the magnetic fields produced by neurons~\cite{BAR2016}, integrated circuits~\cite{NOW2015,HOR2018}, and the nuclear magnetic resonance of fluids~\cite{GLE2018,SMI2019}. 

Diamond magnetometers with larger active volumes are expected to offer the highest sensitivity~\cite{BAR2019}. However, in order to be competitive with existing technologies, they must overcome several technical drawbacks, including high laser-power requirements and poor sensitivity at low frequencies. The most sensitive diamond magnetometer reported to date featured a projected sensitivity of ${\sim}0.9~{\rm pT\,s^{1/2}}$ using $400~{\rm mW}$ of laser power~\cite{WOL2015}. However this magnetometer used a Hahn-echo pulse sequence which limited the bandwidth to a narrow range around 20 kHz. For broadband, low-frequency operation, the highest sensitivity reported to date is ${\sim}15~{\rm pT\,s^{1/2}}$ in the $80\mbox{--}2000~{\rm Hz}$ range, using $\gtrsim 3~{\rm W}$ of laser power~\cite{BAR2016}. A diamond magnetometer based on infrared absorption detection realized a sensitivity of ${\sim}30~{\rm pT\,s^{1/2}}$ at $10\mbox{--}500~{\rm Hz}$, using $0.5~{\rm W}$ of laser power~\cite{CHA2017}.

\begin{figure*}[htpb]

\includegraphics[width=0.93\textwidth]{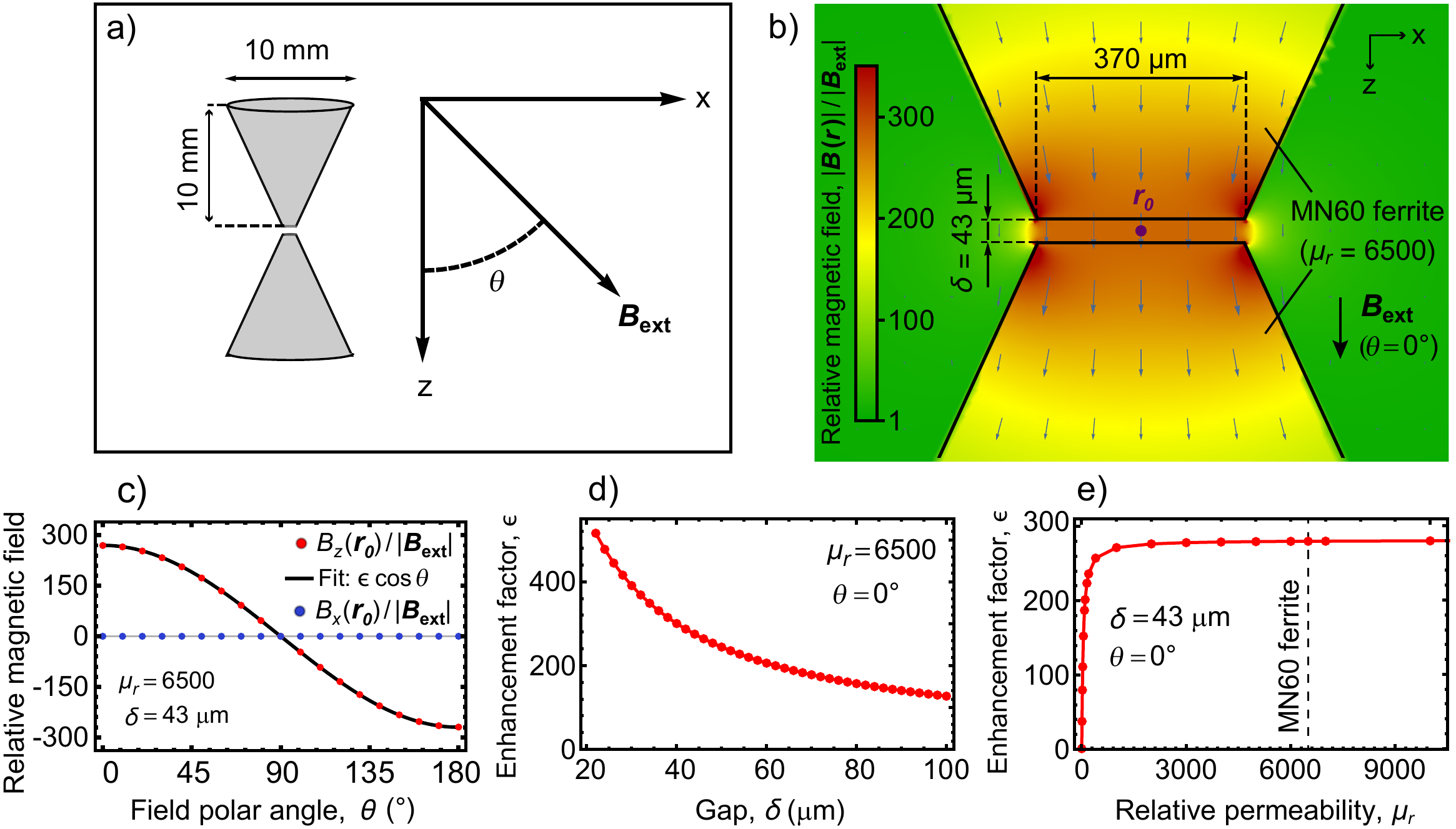}\hfill

\caption{\textbf{Simulations of magnetic flux concentrators.} (a) Model geometry. Two identical solid cones, configured in a bowtie geometry, are placed in an external magnetic field, $\boldsymbol{B_{\rm ext}}$. (b) Simulated x-z plane cut of the relative magnetic field amplitude, $|\boldsymbol{B}(\boldsymbol{r})|/|\boldsymbol{B_{\rm ext}}|$, for cones with relative permeability $\mu_r=6500$ and a tip gap of $\delta=43~{\rm \upmu m}$, upon application of $\boldsymbol{B_{\rm ext}}$ at $\theta=0$. Arrows indicate the direction and magnitude of $\boldsymbol{B}(\boldsymbol{r})$. The point at the geometric center is labeled $\boldsymbol{r_0}$. (c) Vector components of the relative magnetic field amplitude at $\boldsymbol{r_0}$ as a function of $\theta$, for cones with $\mu_r=6500$ and  $\delta=43~{\rm \upmu m}$. The relative axial magnetic field amplitude is fit to the function $B_z(\boldsymbol{r_0})/|\boldsymbol{B_{\rm ext}}|=\epsilon \cos{\theta}$, where in this case $\epsilon=280$. 
		(d) Enhancement factor as a function of $\delta$ for cones with $\mu_r=6500$.   (e) Enhancement factor as a function of $\mu_r$ for $\delta=43~{\rm \upmu m}$. 
		}
\label{fig:sim}
\end{figure*}

To understand the interplay between sensitivity and laser power, we consider a diamond magnetometer based on continuous-wave, fluorescence-detected magnetic resonance (FDMR) spectroscopy. Here, the sensitivity is fundamentally limited by photoelectron shot noise as:
\begin{equation}
\label{eq:psn}\eta_{\rm psn}\approx\frac{\Gamma}{\gamma_{nv}C\sqrt{\xi P_{\rm opt}/E_{ph}}},   
\end{equation}
where $\gamma_{nv}=28~{\rm GHz/T}$ is the NV gyromagnetic ratio, $\Gamma$ is the FDMR full-width-at-half-maximum linewidth, and $C$ is the FDMR amplitude's fractional contrast. The factor $\xi P_{\rm opt}/E_{ph}$ constitutes the photoelectron detection rate, where $P_{\rm opt}$ is the optical excitation power, $\xi$ is the fraction of excitation photons converted to fluorescence photoelectrons, and $E_{ph}=3.7\times10^{-19}~{\rm J}$ is the excitation photon energy ($532~{\rm nm}$). To set an optimistic bound on $\eta_{\rm psn}$, we insert the best reported values ($\xi=0.08$~\cite{WOL2015}, $\Gamma/C=1~{\rm MHz}/0.04$~\cite{BAR2016}) into Eq.~\eqref{eq:psn} to obtain $\eta_{\rm psn}\approx2~{\rm pT\,s^{1/2}\,W^{1/2}}$. Even in this ideal case (\ref{sec:SIshot}), ${\sim}4~{\rm W}$ of optical power is needed to realize a sensitivity of $1~{\rm pT\,s^{1/2}}$, and further improvements become impractical. 

The need for such a high laser power presents challenges for thermal management and has implications for the overall sensor size, weight and cost. Applications which call for sub-picotesla sensitivity, such as magnetoencephalography (MEG)~\cite{BOT2016} and long-range magnetic anomaly detection~\cite{LEN2006,FRO2018}, may require alternative approaches to improve sensitivity. Avenues currently being pursued often focus on reducing the ratio $\Gamma/C$~\cite{BAR2019}. Approaches to reduce $\Gamma$ include lowering $^{13}$C spin density and mitigating strain and electric-field inhomogeneity~\cite{FAN2013,BAU2018}, increasing the nitrogen-to-NV$^-$ conversion yield~\cite{ACO2009,CHA2019,EIC2019}, and designing techniques to decouple NV centers from paramagnetic spins~\cite{DEL2012,BAU2018}. Methods to increase $C$ include using preferentially-aligned NV centers~\cite{OZA2019,OST2019}, detecting infrared absorption~\cite{JEN2014,CHA2017}, and detecting signatures of photo-ionization~\cite{SHI2015,BOU2015,HOP2016}. 

In this Manuscript, we report a complementary approach to improve the sensitivity of diamond magnetometers. Our approach uses microstructured magnetic flux concentrators to amplify the external magnetic field amplitude by a factor of ${\sim}250$ within the diamond sensor. Using a dual-resonance magnetometry technique to suppress the effect of thermal shifts of the NV spin levels, we realize a sensitivity of ${\sim}0.9~{\rm pT\,s^{1/2}}$ in the $10\mbox{--}1000~{\rm Hz}$ range, using a laser power of $200~{\rm mW}$. We show that, with further improvements, a magnetic noise floor of ${\sim}0.02~{\rm pT\,s^{1/2}}$ at $1000~{\rm Hz}$ is possible before ferrite thermal magnetization noise limits the sensitivity.

\section{\label{sec:exp}Experimental design}
Magnetic flux concentrators have previously been used to improve the sensitivity of magnetometers based on the Hall effect~\cite{LER2006}, magnetoresistance~\cite{CAR1998}, magnetic tunnel junctions~\cite{CHA2008}, superconducting quantum interference devices~\cite{BON2002}, and alkali spin precession~\cite{GRI2009}. Typically, the magnetometer is positioned in the gap between a pair of ferromagnetic structures which collect magnetic flux from a larger area and concentrate it into the gap. The fractional increase in magnetic field amplitude due to the flux concentrators, $\epsilon$, is a function of their geometry, gap width, and relative permeability ($\mu_r$). Ideally, the concentrators are formed from a soft magnetic material with low remanence, high $\mu_r$, low relative loss factor~\cite{GRI2009}, and constant susceptibility over a broad range of magnetic field amplitudes and frequencies. The improvement in sensitivity is generally accompanied by a reduction in spatial resolution, as the total magnetometer size is larger (\ref{sec:SIsim}). Diamond sensors usually have sub-mm dimensions, whereas the flux concentrators used here have dimensions of ${\sim}10~{\rm mm}$. Thus our device is best suited for applications that require a spatial resolution ${\gtrsim}10~{\rm mm}$, such as MEG and magnetic anomaly detection. 

\begin{figure*}[ht]
    \centering
\includegraphics[width=0.92\textwidth]{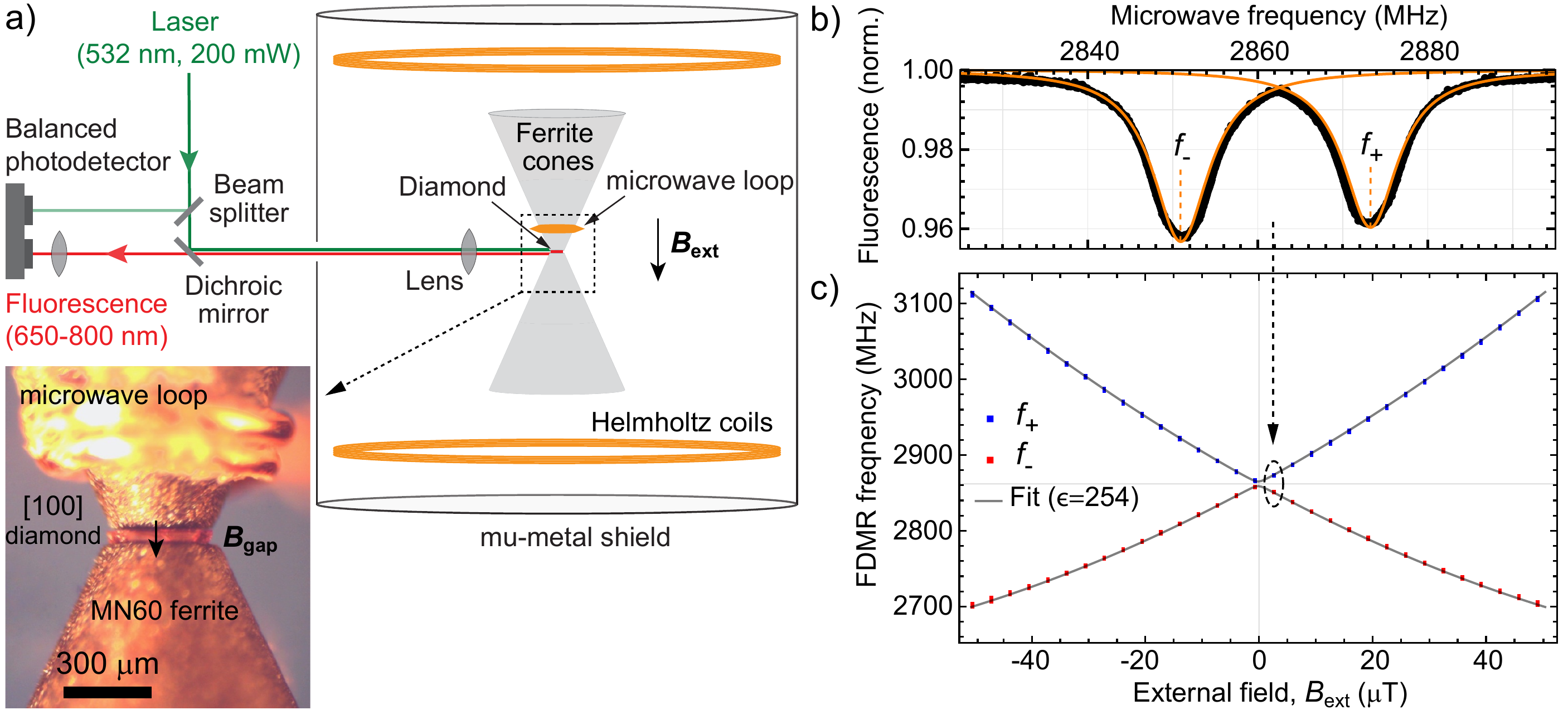}\hfill
\caption{ \textbf{Experimental setup and enhancement measurement.} (a) Schematic of the experimental setup. Inset: photograph of the diamond membrane in the gap between ferrite cones. (b) Fluorescence-detected magnetic resonance (FDMR) spectrum obtained at $B_{\rm ext}=2.62~{\rm \upmu T}$. Two peaks are present, with central frequencies $f_{\pm}$ extracted from Lorentzian fits. (c) Measured FDMR frequencies as a function of $B_{\rm ext}$. Error bars are smaller than the plot markers. The gray solid lines are a fit using the NV spin Hamiltonian (\ref{sec:SIham}), assuming $B_{\rm gap}=\epsilon B_{\rm ext}$, with $\epsilon=254$.}
\label{fig:setup}
\end{figure*}

The optimal flux concentrator geometry depends on a number of factors, which include the sensor dimensions and target application. Here, we consider a pair of identical cones (height: $10~{\rm mm}$, base diameter: $10~{\rm mm}$), with ${\sim}370\mbox{--}{\rm \upmu m}$ diameter flat tips, arranged in a bowtie configuration, Fig.~\ref{fig:sim}(a). A static magnetic field, $\boldsymbol{B_{\rm ext}}$, is applied at an angle $\theta$ from the cone symmetry axis ($\boldsymbol{\hat{z}}$) and the resulting magnetic field, $\boldsymbol{B(r)}$, is simulated using finite-element magnetostatic methods. Figure~\ref{fig:sim}(b) shows a plane-cut of the relative magnetic field amplitude, $|\boldsymbol{B}(\boldsymbol{r})|/|\boldsymbol{B_{\rm ext}}|$, for cones with $\mu_r=6500$ and a tip gap of $\delta=43~{\rm \upmu m}$, upon application of $\boldsymbol{B_{\rm ext}}$ at $\theta=0$. Throughout the gap (\ref{sec:SIsim}), $\boldsymbol{B}(\boldsymbol{r})$ is aligned along $\boldsymbol{\hat{z}}$ with a uniform relative magnetic field $|\boldsymbol{B}(\boldsymbol{r})|/|\boldsymbol{B_{\rm ext}}|\approx280$.

Figure~\ref{fig:sim}(c) shows the vector components of the relative magnetic field at the center of the bowtie geometry ($\boldsymbol{r}=\boldsymbol{r_0}$) as a function of $\theta$. The relative axial magnetic field is well described by $B_z(\boldsymbol{r_0})/|\boldsymbol{B_{\rm ext}}|\approx\epsilon \cos{\theta}$, where $\epsilon$ is the enhancement factor (in this simulation $\epsilon=280$). On the other hand, the relative transverse magnetic field,
$B_x(\boldsymbol{r_0})/|\boldsymbol{B_{\rm ext}}|$, is less than 0.1 for all values of $\theta$. Thus, the structure acts as a filter for the axial component of external magnetic fields, producing a uniform field throughout the gap of:
\begin{equation}
\label{eq:enhancement}
\boldsymbol{B_{\rm gap}}\approx\epsilon\, |\boldsymbol{B_{\rm ext}}|\cos{\theta}\,\boldsymbol{\hat{z}}.
\end{equation}
For the remainder of the manuscript, we consider only external magnetic fields applied along $\boldsymbol{\hat{z}}$ ($\theta=0$) and describe $\boldsymbol{B_{\rm gap}}$ according to Eq.~\eqref{eq:enhancement}.

\begin{figure*}[t]

\includegraphics[width=0.95\columnwidth]{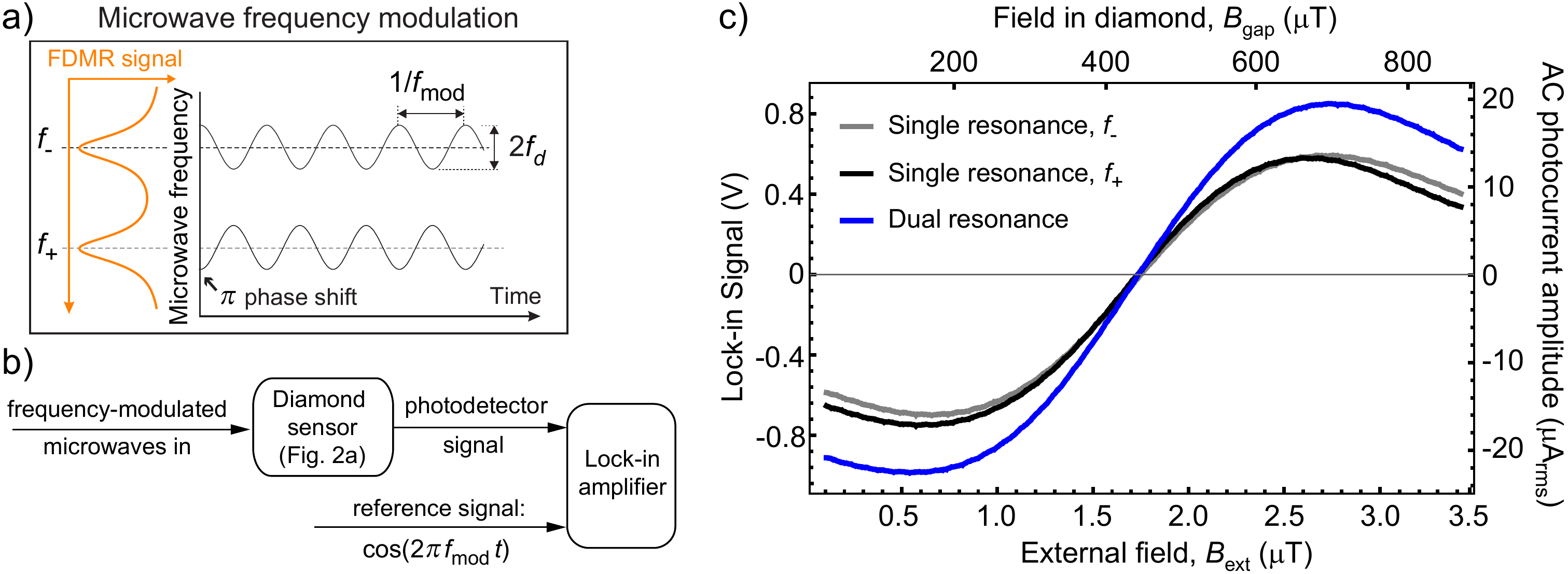} 
\caption{\textbf{Dual-resonance magnetometry concept.} (a) Microwave frequency modulation used for dual-resonance magnetometry. (b) Schematic of the lock-in technique. Both microwave signals depicted in (a) are combined and delivered through the microwave loop. NV fluorescence is continuously excited and its time-varying intensity is recorded by the balanced photodetector. This signal is then fed to a lock-in amplifier and demodulated by the reference signal. (c) Lock-in signal as a function of $B_{\rm ext}$ for both single-resonance and dual-resonance modulation protocols. The microwave frequencies were centered about the $f_{\pm}$ values measured by FDMR spectroscopy at $B_{\rm ext}=1.73~{\rm \upmu T}$. In all cases, $f_{\rm mod}=15~{\rm kHz}$ and the lock-in uses a 12 dB/octave low-pass filter with a $100~{\rm \upmu s}$ time constant. For the $f_-$ scan, the lock-in reference signal had a $\pi$ phase shift relative to the modulation function. The right vertical axis converts the lock-in signal to the amplitude of photocurrent oscillations at $f_{\rm mod}$, which is used to estimate the photoelectron-shot-noise-limited sensitivity, \ref{sec:SIshot}. }
\label{fig:lock}
\end{figure*}

Fig.~\ref{fig:sim}(d) shows simulation results of the enhancement factor as a function of gap length for cones with $\mu_r=6500$. For $\delta$ in the $20\mbox{--}100~{\rm \upmu m}$ range, $\epsilon$ varies from 560 to 120, indicating that large enhancement factors are possible for typical diamond membrane thicknesses. Figure ~\ref{fig:sim}(e) is a plot of the simulated $\epsilon$ as a function of $\mu_r$ for $\delta=43~{\rm \upmu m}$. For $\mu_r\gtrsim500$ the enhancement factor is relatively constant at $\epsilon\approx280$. This indicates that a wide range of magnetic materials can be used for flux concentration and minor variations in $\mu_r$ (due, for example, to temperature variation) have a negligible impact on $\boldsymbol{B_{\rm gap}}$.

We elected to use MN60 ferrite ($\mu_r\approx6500$) as the experimental concentrator material, owing to its low thermal magnetic noise~\cite{GRI2009,KIM2016}. The ferrite cones were micro-machined to have approximately the same dimensions as simulated in Fig.~\ref{fig:sim}. Figure~\ref{fig:setup}(a) depicts the experimental setup. An NV-doped diamond membrane with [100] faces is positioned in the gap between the ferrite cones. The membrane was formed from a commercially-available, type Ib diamond grown by high-pressure high-temperature (HPHT) synthesis. The diamond had been irradiated with $2\mbox{--}{\rm MeV}$ electrons at a dose of ${\sim}10^{19}~{\rm cm^{-2}}$. It was subsequently annealed in a vacuum furnace at $800\mbox{--}1100\degree~{\rm C}$~\cite{FES2019} and mechanically polished and cut into a membrane of dimensions ${\sim}300\times300\times43~{\rm \upmu m}^3$. 

Approximately $200~{\rm mW}$ of light from a $532~{\rm nm}$ laser is focused by a 0.79 NA lens to a ${\sim}40~{\rm \upmu m}$ diameter beam that traverses the diamond membrane parallel to its faces. The same lens is used to collect NV fluorescence, which is then refocused onto one of the channels of a balanced photodetector, producing ${\sim}1.2~{\rm mA}$ of photocurrent. A small portion of laser light is picked off from the excitation path and directed to the other photodetector channel for balanced detection. Microwaves are delivered by a two-turn copper loop wound around one of the ferrite cones. The ferrite cones provide a $\gtrsim2$-fold enhancement in the microwave magnetic field amplitude within the diamond (\ref{sec:SIrabi}). All measurements were performed using ${\lesssim20}~{\rm mW}$ of microwave power.

The ferrite-diamond assembly is positioned at the center of a pair of Helmholtz coils (radius: $38~{\rm mm}$), which produce a homogenous magnetic field parallel to the cone axis of amplitude $B_{\rm ext}$. The coils' current response was calibrated using three different magnetometers (\ref{sec:SIcalib}). A 1.5-mm-thick cylindrical mu-metal shield (diameter: $150~{\rm mm}$, height: $150~{\rm mm}$) surrounds the Helmholtz coils, providing a shielding factor of $\sim100$. 

To measure the enhancement factor, we recorded the NV FDMR spectrum as a function of $B_{\rm ext}$. Figure~\ref{fig:setup}(b) shows a typical FDMR spectrum acquired at $B_{\rm ext}=2.62~{\rm \upmu T}$. Two peaks are present, with central frequencies $f_{\pm}$. These frequencies correspond to NV electron-spin transitions between the $m_s=0$ and $m_s=\pm1$ magnetic sublevels (\ref{sec:SIham}). For magnetic field amplitudes within the diamond in the range $0.5~{\rm mT}\lesssim \epsilon B_{\rm ext}\lesssim5~{\rm mT}$, the transition frequencies may be approximated as:
\begin{equation}
\label{eq:freqs}
f_{\pm}\approx D(\Delta T)\pm \gamma_{nv}\, \epsilon B_{\rm ext}/\sqrt{3},
\end{equation}
where, in our experiments (\ref{sec:SIexpopt}), $D(\Delta T)\approx2862~{\rm MHz}+\chi \Delta T$ is the axial zero-field splitting parameter which shifts with changes in temperature, $\Delta T$, as $\chi\approx-0.1~{\rm MHz/K}$~\cite{TOY2012}. The $1/\sqrt{3}$ factor in Eq.~\eqref{eq:freqs} comes from projecting $\boldsymbol{B_{\rm gap}}$ onto the four NV axes which are all aligned at $55\degree$ with respect to the cone axis. 

Figure~\ref{fig:setup}(c) plots the fitted $f_{\pm}$ values as a function of $B_{\rm ext}$. These data were obtained by scanning $B_{\rm ext}$ back and forth between $\pm50~{\rm \upmu T}$ two times. For a given $B_{\rm ext}$, the extracted $f_{\pm}$ are nearly identical regardless of scan history, indicating negligible hysteresis (\ref{sec:SIhysteresis}). The data were fit according to the NV spin Hamiltonian (\ref{sec:SIham}), which reveals an experimental enhancement factor of $\epsilon=254\pm19$. The uncertainty in $\epsilon$ is primarily due to uncertainty in the $B_{\rm ext}$ current calibration (\ref{sec:SIcalib}). The experimental enhancement factor is ${\sim}10\%$ smaller than the one simulated in Fig.~\ref{fig:sim}(b). This could be explained by a ${\sim}4~{\rm \upmu m}$ increase in $\delta$ due to adhesive between the diamond and ferrite tips (\ref{sec:SIexpcones}).

\begin{figure*}[ht]
    \centering
\includegraphics[width=0.8\textwidth]{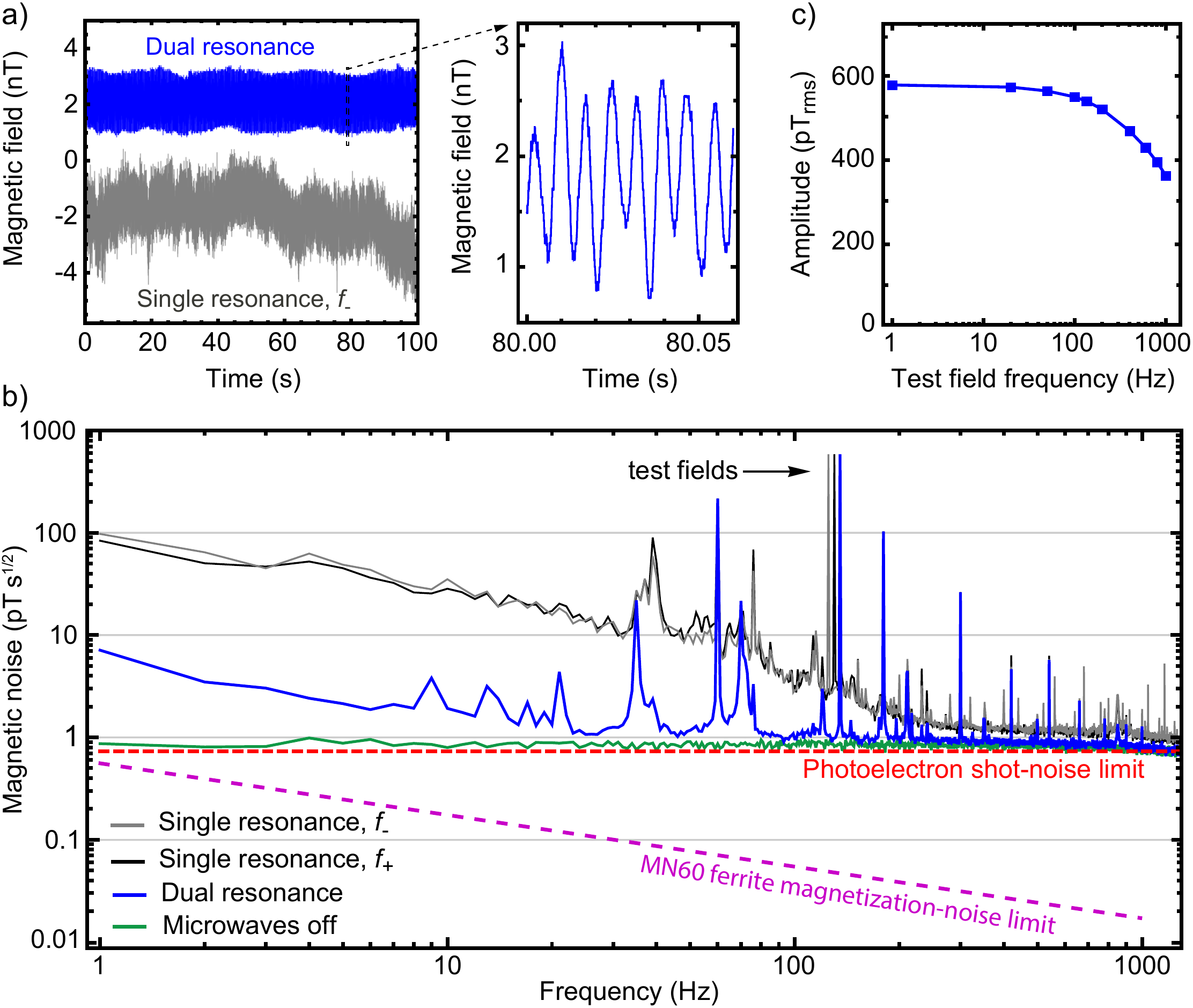}\hfill
\caption{\textbf{Sub-picotesla diamond magnetometry.} (a) Time-domain lock-in signals for single-resonance ($f_-$) and dual-resonance modulation. Throughout, $f_{\rm mod}=15~{\rm kHz}$ and the lock-in uses a 12 dB/octave low-pass filter with a $100~{\rm \upmu s}$ time constant. The adjacent plot is a zoom of the dual-resonance signal where the $580~{\rm pTrms}$ test field at $135~{\rm Hz}$ can be seen. The test-field frequency for $f_+$ and $f_-$ single-resonance experiments were 125 and 130 Hz, respectively, with the same $580~{\rm pTrms}$ amplitude. (b) Magnetic noise spectra of single-resonance (two shades of gray) and dual-resonance (blue) signals. A reference spectrum obtained with microwaves turned off (green) shows noise from the un-modulated photodetector signal. Each spectrum was obtained by dividing a $100\mbox{--}{\rm s}$ data set into one hundred $1\mbox{--}{\rm s}$ segments, taking the absolute value of the Fourier Transform of each segment, and then averaging the Fourier Transforms together. Spectra were normalized such that the test field amplitudes matched the calibrated $580~{\rm pTrms}$ values (\ref{sec:SItest}). The dashed red line is the projected value of $\eta_{\rm psn}$ for dual-resonance magnetometry (\ref{sec:SIshot}). The dashed magenta line is the calculated thermal magnetization noise produced by the ferrite cones (\ref{sec:SInoise}).  (c) Frequency dependence of the test field amplitude measured by dual-resonance magnetometry.}
\label{fig:results}
\end{figure*}

Having established that the ferrite cones provide a ${\sim}250$-fold field enhancement, we now turn to methods of using the device for sensitive magnetometry. A common approach in diamond magnetometry~\cite{SCH2011,SHI2012} is to modulate the microwave frequency about one of the FDMR resonances and demodulate the resulting fluorescence signal using a lock-in amplifier (\ref{sec:SIlockin}). We call this method ``single-resonance'' magnetometry, as each resonance frequency is measured independently. For example, to measure $f_+$, the microwave frequency is varied as $\mathcal{F}(t)\approx f_+ + f_d\cos{(2\pi f_{\rm mod} t)}$, where $f_d$ is the modulation depth and $f_{\rm mod}$ is the modulation frequency. The lock-in amplifier demodulates the photodetector signal using a reference signal proportional to $\cos{(2\pi f_{\rm mod} t)}$. The resulting lock-in output is proportional to variations in $f_+$. 

However, a single FDMR resonance can shift due to changes in \textit{temperature} in addition to magnetic field, see Eq.~\eqref{eq:freqs}. To isolate the shifts due only to changes in magnetic field, the difference frequency $(f_+-f_-)$ must be determined. Previous works accomplished this by measuring both resonances either sequentially~\cite{CLE2015} or simultaneously by multiplexing modulation frequencies~\cite{SCH2018,CLE2018}. The magnetic field was then inferred by measuring $f_+$ and $f_-$ independently and calculating the difference.

Here, we use an alternative ``dual-resonance'' approach, which extracts the magnetic field amplitude directly from a single lock-in measurement (\ref{sec:SIlockin}). Two microwave signal frequencies, centered about $f_{\pm}$, are modulated to provide time-varying frequencies, $\mathcal{F}_{\pm}(t)\approx f_{\pm}\pm\cos{(2\pi f_{\rm mod} t)}$. In other words, each tone is modulated with the same modulation frequency and depth, but with a relative $\pi$ phase shift, Fig.~\ref{fig:lock}(a). The photodetector signal is then demodulated by the lock-in amplifier using a reference signal proportional to $\cos{(2\pi f_{\rm mod} t)}$, Fig.~\ref{fig:lock}(b). In this way, the lock-in output is proportional to $(f_+-f_-)$ and is unaffected by thermal shifts of $D(\Delta T)$. Furthermore, the dual-resonance lock-in signal's response to magnetic fields is larger than in the single-resonance case. Figure~\ref{fig:lock}(c) shows the experimental lock-in signal as a function of $B_{\rm ext}$ for dual-resonance modulation and both of the $f_{\pm}$ single-resonance modulation protocols. The slope for dual-resonance modulation is ${\sim}1.3$ times larger than that of single-resonance modulation. This is close to the expected increase of $4/3$ (\ref{sec:SIdual}).

\section{\label{sec:Measurements} Results}
We next show that the combination of flux concentration and dual-resonance modulation enables diamond magnetometry with sub-${\rm pT\,s^{1/2}}$ sensitivity over a broad frequency range. A $1.73~{\rm \upmu T}$ bias field and $580~{\rm pTrms}$ oscillating test field in the $125\mbox{--}135~{\rm Hz}$ range were applied via the Helmholtz coils. The lock-in signal was continuously recorded for $100~{\rm s}$ using either dual-resonance or single-resonance modulation. Figure~\ref{fig:results}(a) shows the magnetometer signals as a function of time. For single-resonance modulation, the signals undergo low-frequency drifts, likely due to thermal shifts of $D(\Delta T)$. These drifts are largely absent for dual-resonance modulation.

\begin{figure}[t]
    \centering
\includegraphics[width=\columnwidth]{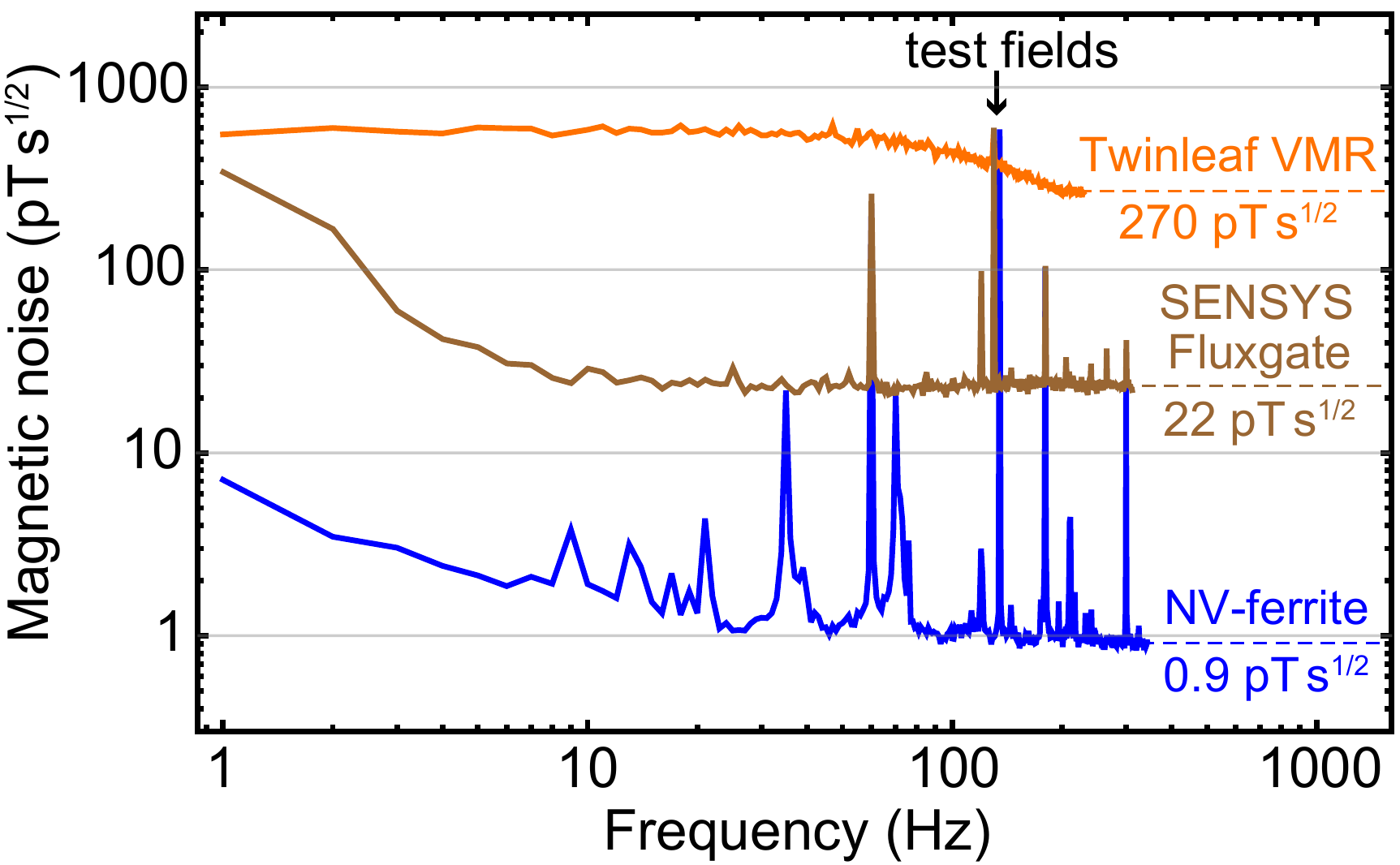}\hfill
\caption{\textbf{Magnetometer comparison.} Magnetic noise spectra of a commercial magnetoresistive magnetometer (Twinleaf VMR), fluxgate magnetometer (SENSYS FGM-100) and our dual-resonance NV-ferrite magnetometer reproduced from Fig.~\ref{fig:results}(b). Each magnetometer was placed in a similar location within the experimental apparatus and subject to the same bias and test field amplitudes. The test field frequency was $130~{\rm Hz}$ for both commercial sensors and $135~{\rm Hz}$ for NV-ferrite. The manufacturer-specified sensitivities are $300~{\rm pT/ \sqrt{Hz}}$ and $10~{\rm pT/\sqrt{Hz}}$ for the VMR and fluxgate, respectively.}
\label{fig:mags}
\end{figure}

Figure~\ref{fig:results}(b) shows the magnetic noise spectrum for the different modulation techniques. In addition to the calibrated test field signals, numerous peaks appear for both single and dual-resonance modulation. We attribute these peaks to ambient magnetic noise that is not sufficiently attenuated by the single-layer mu-metal shield. In regions without peaks, the noise floor for single-resonance magnetometry is ${\sim}1.5~{\rm pT\,s^{1/2}}$ for frequencies ${\gtrsim}300~{\rm Hz}$, but it exhibits nearly $1/f$ behavior for lower frequencies. On the other hand, the noise floor for dual-resonance magnetometry is ${\sim}0.9~{\rm pT\,s^{1/2}}$ for frequencies ${\gtrsim}100~{\rm Hz}$ and remains at this level, to within a factor of two, for frequencies down to ${\sim}10~{\rm Hz}$. The remaining noise below $10~{\rm Hz}$ may be due to thermal variation in the gap length, $\delta$ (\ref{sec:SIgap}). For reference, a spectrum obtained with the microwaves turned off is also shown. It features a constant noise floor of ${\sim}0.8~{\rm pT\,s^{1/2}}$ throughout the $1\mbox{--}1000~{\rm Hz}$ frequency range. This level is consistent with the projected photoelectron shot-noise limit, $\eta_{\rm psn}=0.72~{\rm pT\,s^{1/2}}$, which was calculated based on the average photocurrent and lock-in slope (\ref{sec:SIshot}). 

The frequency response of the magnetometer was determined by recording magnetic spectra at different test-field frequencies, while holding the amplitude of the driving current constant. Figure~\ref{fig:results}(c) plots the test-field amplitude, recorded by dual-resonance diamond magnetometry, as a function of frequency. The amplitude decays by less than a factor of two over the $1\mbox{--}1000~{\rm Hz}$ range. The observed decay is due to a combination of the lock-in amplifier's low-pass filter and a frequency-dependent magnetic field attenuation due to metal components within the Helmholtz coils (\ref{sec:SItest}).

Finally, we compared the performance of our magnetometer with two commercial vector sensors: a magnetoresistive magnetometer and a fluxgate magnetometer. Figure~\ref{fig:mags} shows the magnetic noise spectra obtained under comparable experimental conditions.  Evidently, the NV-ferrite magnetometer outperforms the commercial magnetometers throughout the frequency range.

\section{\label{sec:Discussions}Discussion and conclusion}
The demonstration of broadband, sub-picotesla diamond magnetometry is a significant step towards applications in precision navigation, geoscience, and medical imaging. Since only $200~{\rm mW}$ of laser power and $20~{\rm mW}$ of microwave power were used, the device holds promise for future miniaturization and parallelization efforts. Moreover, our magnetometer operates at microtesla ambient fields, which raises the intriguing possibility of operating in Earth's magnetic field without an additional bias field.

Our implementation used a commercially-available, type Ib HPHT diamond processed using standard electron-irradiation and annealing treatments~\cite{ACO2009}. This material exhibits relatively broad FDMR resonances ($\Gamma\approx9~{\rm MHz}$), which leads to a photoelectron-shot-noise-limited sensitivity of $\eta_{\rm psn}=0.72~{\rm pT\,s^{1/2}}$ even after the ${\sim}250$-fold flux-concentrator field enhancement. State-of-the-art synthetic diamonds have recently been fabricated that feature several orders of magnitude narrower resonances~\cite{BAU2018,ZHE2019}. The excitation photon-to-photoelectron conversion efficiency in our experiments ($\xi\approx10^{-2}$) could also be improved by at least an order of magnitude with optimized collection optics~\cite{WOL2015}. With these additions, $\eta_{\rm psn}$ could be further improved by several orders of magnitude, Eq.~\eqref{eq:psn}. However, at this level, thermal magnetization noise intrinsic to the flux concentrators becomes relevant.

Thermal magnetic noise originating from dissipative materials can be estimated using fluctuation-dissipation methods~\cite{GRI2009,LEE2008}. The noise has contributions due to thermal eddy currents and magnetic domain fluctuations. As discussed in~\ref{sec:SInoise}, we find that thermal eddy currents in the ferrite cones produce an effective white magnetic noise of ${\sim}7\times10^{-5}~{\rm pT\,s^{1/2}}$. This negligibly-low noise level is a consequence of our choice of low-conductivity ferrite. On the other hand, thermal magnetization noise results in a larger, frequency-dependent magnetic noise. At 1 Hz, this noise is $0.5~{\rm pT\,s^{1/2}}$, and it scales with frequency as $f^{-1/2}$, reaching ${\sim}0.02~{\rm pT\,s^{1/2}}$ at $1~{\rm kHz}$. This noise, shown in Fig.~\ref{fig:results}(b), is not a limiting factor in our experiments, but it may have implications for future optimization efforts. If a material with a lower relative loss factor could be identified, it would result in lower thermal magnetization noise (\ref{sec:SInoiseM}).

In summary, we have demonstrated a diamond magnetometer with a sensitivity of ${\sim}0.9~{\rm pT\,s^{1/2}}$ over the $10\mbox{--}1000~
{\rm Hz}$ frequency range. The magnetometer operates at ambient temperature and uses $0.2~{\rm W}$ of laser power. These improved sensor properties are enabled by the use of ferrite flux concentrators to amplify magnetic fields within the diamond sensor. Our results may be immediately relevant to applications in precision navigation, geoscience, and medical imaging. More broadly, the use of micro-structured magnetic materials to manipulate magnetic fields offers a new dimension for diamond quantum sensors, with potential applications in magnetic microscopy~\cite{LOU2019,STE2013,GLE2015,FES2019,MCC2019,BAR2016,NOW2015,HOR2018} and tests of fundamental physics~\cite{CHU2016}.

\begin{acknowledgments}
The authors acknowledge advice and support from A. Laraoui, Z. Sun, D. Budker, P. Schwindt, A. Mounce, M. S. Ziabari, B. Richards, Y. Silani, F. Hubert, and M. D. Aiello. This work was funded by NIH grants 1R01EB025703-01 and 1R21EB027405-01, NSF grant DMR1809800, and a Beckman Young Investigator award.

\textbf{Competing interests}
I. Fescenko, A. Jarmola, and V. M. Acosta are co-inventors on a pending patent application. A. Jarmola is a co-founder of ODMR Technologies and has financial interests in the firm. The remaining authors declare no competing financial interests.

\textbf{Author contributions}
V. M. Acosta and I. Savukov conceived the idea for this study in consultation with I. Fescenko and A. Jarmola. I. Fescenko carried out simulations, performed experiments, and analyzed the data with guidance from V. M. Acosta. P. Kehayias, J. Smits, J. Damron, N. Ristoff, N. Mosavian, and A. Jarmola contributed to experimental design and data analysis. All authors discussed results and helped write the paper.

\end{acknowledgments}

\clearpage

\appendix

\setcounter{equation}{0}
\setcounter{section}{0}

\setcounter{table}{0}

\makeatletter
\renewcommand{\thetable}{A\arabic{table}}

\renewcommand{\theequation}{A\Roman{section}-\arabic{equation}}
\renewcommand{\thefigure}{A\arabic{figure}}
\renewcommand{\thesection}{Appendix~\Roman{section}}

\makeatother

\begin{center}
\section{\label{sec:SIham}} 
\setlength{\parskip}{-0.8em}{
\textbf {NV electron spin Hamiltonian}}
\end{center}

Neglecting hyperfine coupling (which is not resolved in our experiments), the NV ground-state electron spin Hamiltonian can be written as~\cite{BudkerTut}:
 \begin{equation}
\label{eq:ham}\frac{{\hat H}}{h}=D S^2_{z'}+E(S^2_{x'}-S^2_{y'})+\gamma_{nv}\pmb{B}\cdot\pmb{S},
 \end{equation}
where $h$ is Planck's constant, $\gamma_{nv}=28.03~{\rm GHz/T}$ is the NV gyromagnetic ratio, and $E\approx3~{\rm MHz}$ is the transverse zero-field splitting parameter. The axial zero-field splitting parameter, $D\approx2862~{\rm MHz}$, is temperature dependent, as discussed in \ref{sec:SIexpopt}. ${\pmb S}=(S_{x'},S_{y'},S_{z'})$ are dimensionless electron spin operators, and the ${\pmb z'}$ direction is parallel to the NV symmetry axis. For a magnetic field of amplitude $B_{\rm gap}$ applied normal to a diamond with [100] faces, the Hamiltonian for NV centers aligned along any of the four possible axes is the same. In matrix form, it is: 
\newcommand\scalemath[2]{\scalebox{#1}{\mbox{\ensuremath{\displaystyle #2}}}}
\arraycolsep=6pt\def\arraystretch{2}
\begin{equation}
\label{eq:htot}
\frac{{\hat H}}{h}=\left(
\scalemath{1}{
\begin{array}{lcr}
D+\frac{\gamma_{nv}B_{\rm gap}} {\sqrt3}  &  \frac{\gamma_{nv}B_{\rm gap}}{\sqrt3} & E\\
\frac{\gamma_{nv}B_{\rm gap}}{\sqrt3}   &  0  &   \frac{\gamma_{nv}B_{\rm gap}}{\sqrt3}\\
E & \frac{\gamma_{nv}B_{\rm gap}}{\sqrt3} &  D-\frac{\gamma_{nv}B_{\rm gap}}{\sqrt3}\\
\end{array}}
\right),
\end{equation}
The eigenstates and eigenfrequencies can be found by diagonalizing the Hamiltonian. The two microwave transition frequencies observed in our experiments, $f_{\pm}$, correspond to the frequency differences between the eigenstate with largely $m_s=0$ character and the eigenstates with largely $m_s=\pm1$ character. We used this Hamiltonian to fit the $f_{\pm}$ versus $B_{\rm ext}$ data in Fig.~\ref{fig:setup}(c). We assumed $B_{\rm gap}=\epsilon B_{\rm ext}$ and used solutions to Eq.~\ref{eq:htot} to fit for $\epsilon=254$. The values of $E$ and $D$ were determined separately from low-field FDMR data and were not fit parameters. 

Note that Eq.~\eqref{eq:freqs} in the main text, which approximates $f_\pm$ as being linearly dependent on $B_{\rm ext}$, is merely a convenient approximation. As can be seen in Fig.~\ref{fig:setup}(c), the exact values of $f_{\pm}$ are generally nonlinear functions of $B_{\rm ext}$. This is especially pronounced near zero field, $\epsilon\,|B_{\rm ext}|\lesssim E/\gamma_{nv}\approx0.1~{\rm mT}$, where $f_{\pm}$ undergo an avoided crossing, and also at high field, where mixing due to transverse fields produces nonlinear dependence. However, for magnetic fields $0.5~{\rm mT}\lesssim \epsilon B_{\rm ext}\lesssim5~{\rm mT}$, the transition frequencies $f_{\pm}$ are approximately linear in $B_{\rm ext}$.

\begin{center}
\section{\label{sec:SIsim}} 
\setlength{\parskip}{-0.8em}{
\textbf {Flux concentrator simulations}}
\end{center}

\begin{figure}[t]
    \centering
\includegraphics[width=\textwidth]{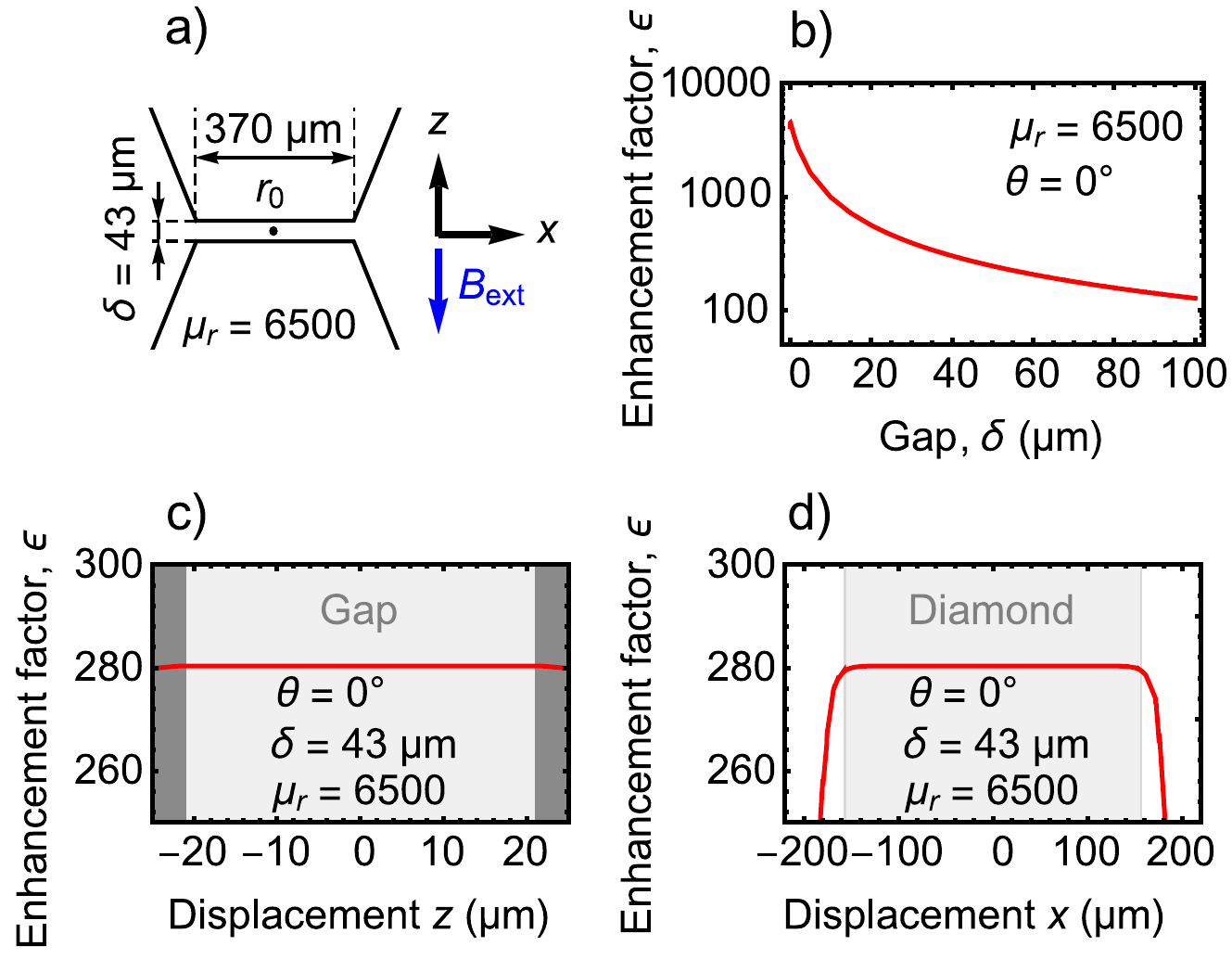}\hfill
       \caption{
\textbf{Enhancement factor and field homogeneity.}  
    (a) The model geometry. See Fig.~\ref{fig:sim}(a) for additional dimensions.
    (b) Enhancement factor, $\epsilon$, as a function of the gap length, $\delta$. 
    (c) Enhancement factor as a function of the axial displacement $z$. The gap is shaded in light gray, while the ferrite concentrators are shaded in dark gray.  (d) Enhancement factor as a function of the transverse displacement $x$.}
    \label{fig:cuts}
\end{figure}

Our flux concentrator model and simulations are described in Sec.~\ref{sec:exp} and Fig.~\ref{fig:sim} of the main text. Here we describe supplementary results demonstrating the field homogeneity in the gap, the enhancement factor as the gap length approaches zero, and the approximate point spread function. Figure~\ref{fig:cuts}(a) describes the geometry used for the simulations. Figure~\ref{fig:cuts}(b) shows the enhancement factor as a function of $\delta$, with the range extending to $\delta\approx0$. The largest enhancement factors are observed for small gaps, approaching $\epsilon=5000$ for $\delta=0$. We chose a gap of $\delta\approx43~{\rm \upmu m}$ in our experiments as a compromise that offers moderate enhancement ($\epsilon\approx250$) while still providing substantial optical access and straightforward fabrication and construction. 

To visualize the homogeneity of the magnetic field within the gap, we plot line cuts of the relative field amplitude along the axial and transverse directions. Figure~\ref{fig:cuts}(c) shows the relative magnetic field along the cone symmetry axis. Figure~\ref{fig:cuts}(d) shows the relative field along a transverse line passing through $\boldsymbol{r_0}$. Both plots predict a high degree of magnetic field homogeneity; residual variations of the relative field are $\lesssim1\%$ throughout the region filled by the diamond membrane.

\begin{figure}[htb]
    \centering
\includegraphics[width=0.95\columnwidth]{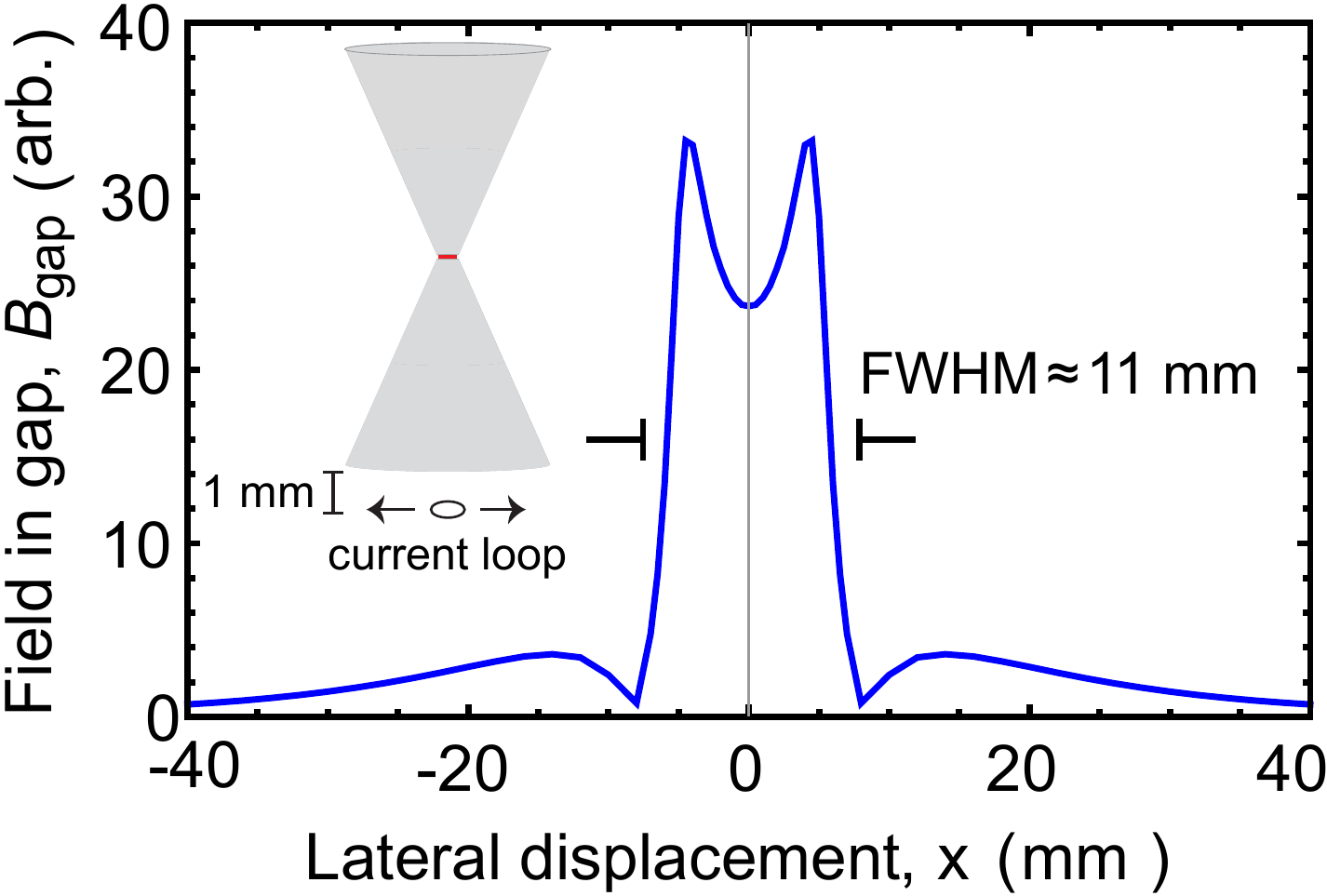}\hfill
       \caption{
\textbf{Flux concentrator point spread function.}  The value of $B_{\rm gap}$ due to a small current loop located below the device is recorded as a function of lateral displacement. The resulting field profile has a FWHM linewidth of ${\sim}11~{\rm mm}$. Inset: geometry for scanning. }
\label{fig:psf}
\end{figure}

Future NV-flux concentrator devices may involve the use of sensor arrays to perform imaging. While a detailed analysis of the design space for imaging applications is beyond the scope of this work, we performed simulations to estimate the point spread function of our device. A small ($1\mbox{--}{\rm mm}$ diameter) current loop was positioned to have an axial displacement of $1~{\rm mm}$ below the base of the bottom cone. The magnetic field amplitude in the gap, $B_{\rm gap}$, was simulated as a function of the current loop's lateral displacement, $x$. Figure~\ref{fig:psf} shows the resulting magnetic field profile. While it does not a have simple Gaussian shape, it can be approximated as having a full-width-at-half-maximum (FWHM) resolution of ${\sim}11~{\rm mm}$.

\begin{center}
\section{\label{sec:SIexpcones}} 
\setlength{\parskip}{-0.8em}{
\textbf {Experimental setup: cones}}
\end{center}

The ferrite cones were ordered from Precision Ferrites \& Ceramics, Inc. The diamond membrane was glued on the tip of one of the ferrite cones with LOCTITE AA3494 UV-curing adhesive. The second cone with the microwave loop was mounted inside a metallic holder and micro-positioned to contact the exposed face of the diamond membrane by use of a Thorlabs MicroBlock Compact Flexure Stage MBT616D. When in the desired position, the holder was glued to the support of the bottom cone by superglue, and then detached from the micro-positioning stage. 

\begin{figure}[htb]
    \centering
\includegraphics[width=0.95\columnwidth]{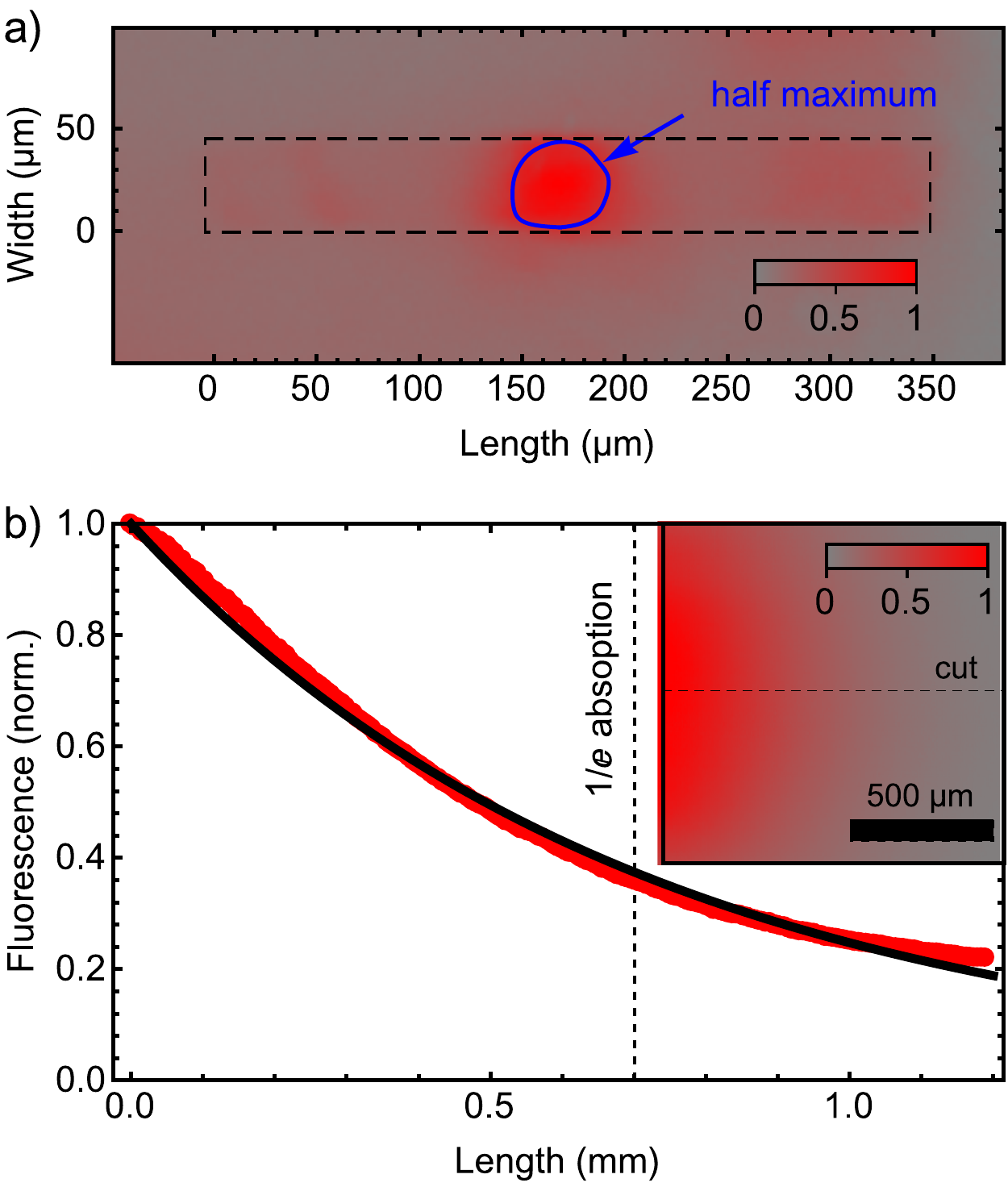}\hfill
       \caption{
\textbf{Beam profile and absorption length.}  (a) Image of the fluorescence spot at the entrance edge of the diamond membrane. The FWHM spot diameter is $\sim40~{\rm \upmu m}$. The dashed lines indicate the approximate edges of the diamond. (b) Fluorescence intensity produced by a ${\sim}1~{\rm mm}$ diameter laser beam entering the edge of a diamond membrane. The inset shows a fluorescence image of the top face. Red markers depict the normalized fluorescence intensity along the cut shown by the dashed line in the inset. The black solid line is an exponential fit, revealing a $1/e$ absorption length of $0.6~{\rm mm}$.}
\label{fig:laser}
\end{figure}

\begin{figure*}[ht]
    \centering
\includegraphics[width=\textwidth]{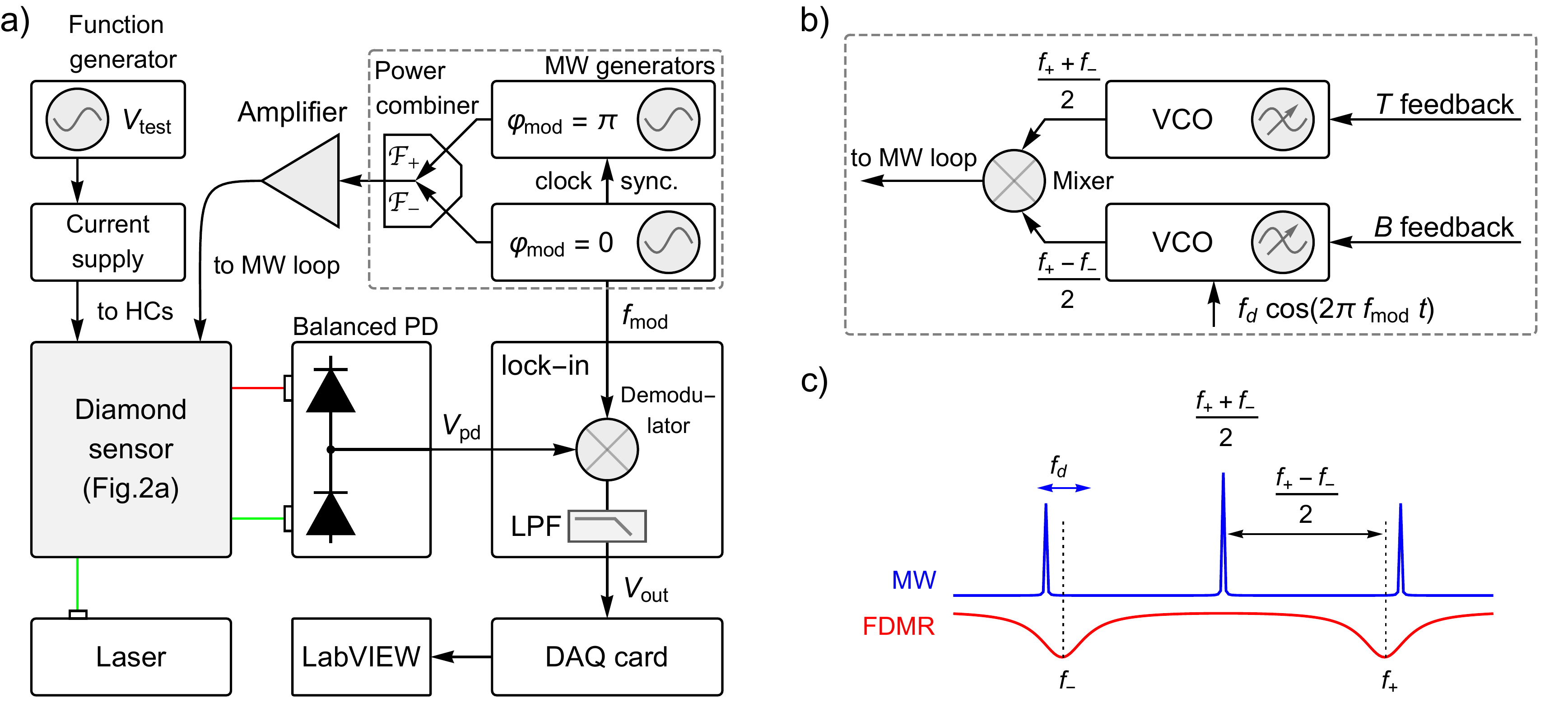}\hfill
       \caption{
\textbf{Electronics.} (a) Schematic of the electronics portion of the experimental apparatus. $V_{\rm pd}$ is the photodetector signal, $V_{\rm out}$ is the lock-in amplifier's in-phase output signal, $V_{\rm test}$ is the test signal waveform, and $f_{\rm mod}$ is the modulation frequency. (b) Alternative electronic scheme for dual-resonance microwave signal generation and feedback. A voltage-controlled oscillator (VCO) produces a carrier frequency $f_{\rm car}=(f_++f_-)/2\approx D(\Delta T)$ that is mixed with the signal from a second VCO with frequency $f_{\rm diff}=(f_+-f_-)/2$, creating two sidebands at the FDMR frequencies. The sidedand frequencies are modulated by adding a reference signal $f_{\rm d}\cos{(2\pi f_{\rm mod} t)}$ to the second VCO. This arrangement allows for rapid feedback to correct for temperature and magnetic field drifts by adjusting the bias voltage to the VCOs. (c) Microwave signal spectrum resulting from the alternative electronics scheme in (b). A typical FDMR spectrum is shown in red for reference.
DAQ: data acquisition card;
HC: Helmholtz coils;
LPF: low-pass filter;
MW: microwave;
PD: photodetector. }
\label{fig:electronics}
\end{figure*}

\begin{center}
\section{\label{sec:SIexpopt}} 
\setlength{\parskip}{-0.8em}{
\textbf {Experimental setup: optics}}
\end{center}

\begin{figure}[hb]
    \centering
\includegraphics[width=0.75\columnwidth]{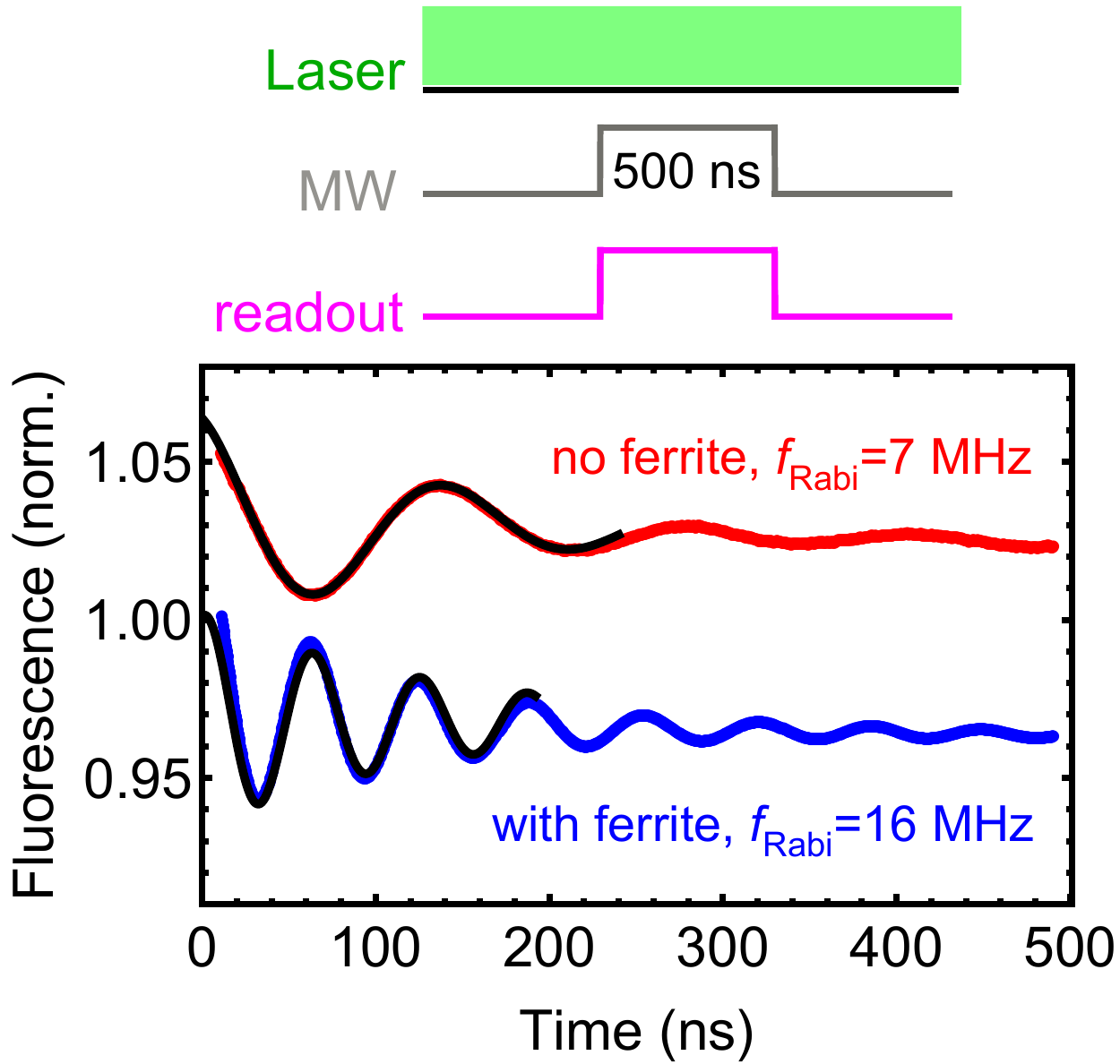}\hfill
       \caption{\label{fig:rabi}
\textbf{Rabi frequency with and without ferrite.}  
(top) Protocol used to observe continuous-wave Rabi oscillations. (bottom) Rabi oscillations observed with and without ferrite cones (the setup was identical otherwise). Black solid curves are fits to an exponentially-damped sinusoidal function revealing $f_{Rabi}=16~{\rm MHz}$ with ferrite and $f_{Rabi}=7~{\rm MHz}$ without ferrite. A microwave power of ${\sim}10~{\rm W}$ was used for both traces in order to clearly visualize the Rabi oscillations.}
\end{figure}

To excite NV fluorescence, a Lighthouse Photonics Sprout-G laser is used to form a collimated beam of 532 nm light. The beam is focused with a Thorlabs aspheric condenser lens ACL25416U-B (NA=0.79) onto the edge of the diamond membrane. Fluorescence is collected by the same lens and is spectrally filtered by a Semrock FF560-FDi01-25x36 dichroic mirror. A second lens re-images the fluorescence onto a photodetector. For magnetometry experiments, including all data in the figures in the main text, we used a Thorlabs PDB210A balanced photodetector. For beam characterization (Fig.~\ref{fig:laser}), we used a CMOS image sensor, and for observing Rabi oscillations (Fig.~\ref{fig:rabi}), we used a Thorlabs PDA8A high-speed photodetector. Figure~\ref{fig:laser}(a) shows an image of the fluorescence spot from the entrance edge of the diamond membrane. The FWHM spot diameter of ${\sim}40~{\rm \upmu m}$ was selected to match the diamond membrane thickness. It was adjusted by tailoring additional telescoping lenses in the excitation path.

With this optical system, we obtained a excitation photon-to-photoelectron conversion efficiency of $\xi\approx0.01$. The primary factors limiting $\xi$ are due to the limited optical access afforded by the ferrite cones, loss of fluorescence exiting orthogonal faces of the diamond membrane, and incomplete absorption of the excitation beam within the diamond. To characterize the latter, we used a separate apparatus to image the fluorescence from the top face of a larger membrane, Fig.~\ref{fig:laser}(b). This larger membrane was the starting piece from which we cut the smaller membrane used in magnetometry experiments. We found that the $1/e$ absorption length of this material is $0.6~{\rm mm}$. Thus we expect that only ${\sim}40\%$ of the laser light was absorbed in the ${\sim}300~{\rm \upmu m}$-long diamond membrane used in magnetometry experiments. This approximation neglects the effects of laser light that is reflected at the air-diamond interfaces.
 
 The large absorbed optical power results in significant heating of the diamond membrane. The experimentally-measured axial zero-field splitting parameter $D\approx2862~{\rm MHz}$, Fig.~\ref{fig:setup}(c), indicates a local diamond temperature of ${\sim}385~{\rm K}$~\cite{TOY2012}. While the elevated temperature leads to a large shift in $D$, it does not significantly diminish the contrast or broaden the FDMR resonances. Future devices may employ active cooling or optimized heat sinks to reduce the diamond temperature.

\begin{center}
\section{\label{sec:SIexpelec}} 
\setlength{\parskip}{-0.8em}{
\textbf {Experimental setup: electronics}}
\end{center}

Figure~\ref{fig:electronics}(a) shows a schematic of the electronic devices used in our experimental setup. Microwaves are supplied by two Stanford Research SG384 signal generators. The clocks of the generators are synchronized by passing the 10 MHz frequency reference output of one generator to the frequency reference input of the other. Both generators are configured to modulate the microwave frequency with a modulation frequency $f_{\rm mod}=15$~kHz and depth $f_d=3.3~{\rm MHz}$. In dual-resonance modulation, the signal generators are configured such that their modulation functions, $\mathcal{F}_{\pm}$, have a relative $\pi$ phase shift (see Sec.~\ref{sec:exp}). The signals from both generators are combined with a Mini-Circuits ZAPD-30-S+ 2-way power combiner, amplified by a Mini-Circuits amplifier ZHL-16W-43-S+, and finally delivered to a two-turn microwave loop made from polyurethane-enameled copper wire (38~AWG). Prior to performing dual-resonance magnetometry, the microwave powers for each $f_{\pm}$ resonance were independently adjusted to give approximately the same lock-in slope  Fig.~\ref{fig:lock}(c).

The photodetector output signal, $V_{\rm pd}$, is fed to a Signal Recovery 7280 lock-in amplifier using $50~\Omega$ termination. The lock-in multiplies $V_{\rm pd}$ by a reference signal, proportional to $\cos{(2\pi f_{\rm mod} t)}$, output from one of the signal generators. The demodulated signal is processed by the lock-in's low pass filter, which was set to 12 dB/octave with a $100~\upmu$s time-constant. The lock-in amplifier's in-phase component, $V_{\rm out}$, is digitized at 50 kS/s by a National Instrument USB-6361 data acquisition unit. 

External fields, $B_{\rm ext}$, are produced by a pair of Helmholtz coils (radius: $38~{\rm mm}$) driven by a Twinleaf CSUA-50 current source. To create oscillating test signals, a Teledyne LeCroy WaveStation 2012 function generator provides a sinusoidal waveform, $V_{\rm test}$, to the modulation input of the current source. The same function generator was used to slowly sweep the magnetic field for the lock-in signals shown in  Fig.~\ref{fig:lock}c (in this case, no oscillating test signals were applied).

While our tabletop prototype uses scientific-grade microwave generators, a simpler system could be used to deliver the requisite dual-resonance microwave waveforms. Figures~\ref{fig:electronics}(b) shows an alternative scheme which uses only voltage-controlled oscillators and a mixer. This scheme has the benefit of allowing for rapid feedback to compensate for thermal and magnetic field drifts, which would enable a higher dynamic range~\cite{CLE2018}.

\begin{center}
\section{\label{sec:SIlockin}} 
\setlength{\parskip}{-0.8em}{
\textbf {Dual-resonance magnetometry}}
\end{center}

We perform our magnetometry experiments with a lock-in amplifier in order to reduce technical noise, particularly at low frequencies. Such noise could arise from a variety of sources, but a common source in NV magnetometry experiments is due to intensity fluctuations of the laser that are not fully canceled by balanced photodetection. The lock-in method allows us to tune our photodetector signal to a narrow frequency band, where such technical noise is minimal. In our experiments, this is accomplished by modulating the microwave frequency at a modulation frequency $f_{\rm mod}=15~{\rm kHz}$ and depth $f_d=3.3~{\rm MHz}$. The resulting photodetector signal, $V_{\rm pd}$, has components at $f_{\rm mod}$ and higher harmonics, in addition to the DC level. The lock-in amplifier isolates the component at $f_{\rm mod}$, in a phase-sensitive manner, by multiplying $V_{\rm pd}$ by a reference signal proportional to $\cos{(2\pi f_{\rm mod} t)}$. The product signal is passed through a low-pass filter, and the in-phase component, $V_{\rm out}$, serves as the magnetometer signal. 

The lock-in signal, $V_{\rm out}$, can be converted to magnetic field units by one of two methods. In the first case, one can sweep the magnetic field and measure the dependence of $V_{\rm out}$ on $B_{\rm ext}$, as in Fig.~\ref{fig:lock}(c) of the main text. The slope can be used to infer the conversion of $V_{\rm out}$ to magnetic field units. This method works well provided that the slope never changes. In practice, the slope can change due to drifts of the laser or microwave powers. It also can't account for any dependence of $V_{\rm out}$ on magnetic field frequency, as the slope is measured at DC. Thus, we always apply a calibrated oscillating test field and re-normalize our magnetometer conversion based on the observed amplitude. Typically the difference in conversion factors using the two methods is small ($\lesssim10\%$).

We now turn to describing the principle of dual-resonance magnetometry. In single-resonance magnetometry, the microwave frequency is modulated about one of the FDMR resonances (for example, $f_+)$ and demodulated at the same frequency. The in-phase lock-in output $V_{\rm out}$ is proportional to small deviations in $f_+$. This allows one to infer both the magnitude and sign of changes in $f_+$. If the relative phase between the microwave modulation function, $\mathcal{F}_+$, and the reference signal were shifted by $\pi$ radians, the magnitude of $V_{\rm out}$ would be the same but the sign would reverse.

In dual-resonance magnetometry, we exploit this feature of phase-sensitive detection. The microwave modulation function for one resonance has a $\pi$ phase shift with respect to the modulation function of the second resonance. The reference signal has the phase of the first modulation function. In this way, if both $f_+$ and $f_-$ shift by equal amounts in the same direction [due to a change in $D(\Delta T)$], their contributions to the lock-in signal cancel and $V_{\rm out}=0$. If $f_+$ and $f_-$ shift by equal amounts but in opposite directions (due to a change in $B_{\rm ext}$), their contributions to the lock-in signal add together and $V_{\rm out}$ changes in proportion to their shift. In other words, the lock-in output is unaffected by thermal shifts of the NV spin levels (which shift $f_+$ and $f_-$ by equal amounts in the same direction), but it remains proportional to changes in magnetic field (which shift $f_+$ and $f_-$ by approximately equal amounts in opposite directions).

Note that dual-resonance modulation could also be used to make an NV thermometer which is unaffected by changes in magnetic field. This would be accomplished by applying the same modulation phase to both $\mathcal{F}_{\pm}$ signals and monitoring the in-phase lock-in signal.

\begin{center}
\section{\label{sec:SIdual}} 
\setlength{\parskip}{-0.8em}{
\textbf {Sensitivity enhancement in dual-resonance magnetometry}}
\end{center}

The dual-resonance magnetometry approach was primarily used because it is unaffected by thermal shifts of the NV spin levels. This enabled better low-frequency performance. However the dual-resonance approach also has a fundamental advantage in sensitivity for all frequencies. Compared to the single-resonance approach, it offers a ${\sim}4/3$-fold improvement in photoelectron-shot-noise-limited sensitivity. This improvement comes about due to a  ${\sim}4/3$-fold increase in the FDMR contrast.

\begin{figure}[t]
    \centering
\includegraphics[width=0.95\columnwidth]{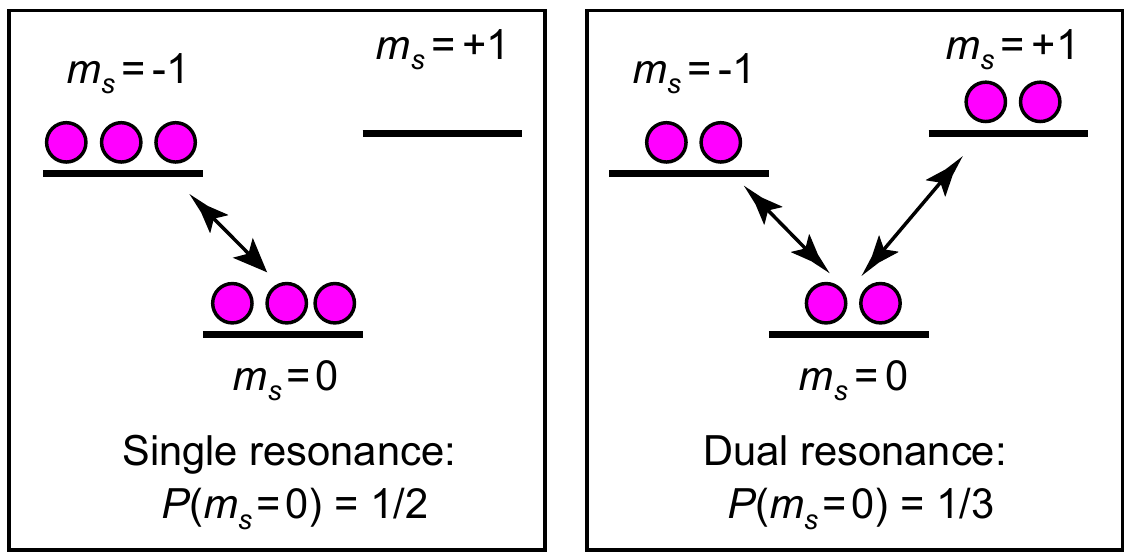}\hfill
       \caption{\label{fig:levels}
\textbf{Single and dual-resonance spin populations.}  NV spin level populations, represented by the number of magenta circles, are shown under single-resonance and dual-resonance microwave excitation.}
\end{figure}

To understand where the factor of $4/3$ arises, consider the limiting case when the microwave excitation rate is much larger than the optical excitation rate. In this regime, a resonant microwave field drives the spin levels it interacts with into a fully mixed state, Fig.~\ref{fig:levels}. For single-resonance excitation, when the microwave field is on resonance, the probability that NV centers will be in the $m_s=0$ level is $P_0=1/2$. For dual-resonance excitation, both microwave transitions share the $m_s=0$ level and thus $P_0=1/3$ when both microwave fields are on resonance. Defining the fluorescence intensity of an NV center in the $m_s=0$ level as $I_0$ and the fluorescence intensity of an NV center in either of the $m_s=\pm1$ levels as $I_1$, the FDMR contrast is given by:
\begin{equation}
\label{eq:intensity}
C=\frac{I_0-[P_0I_0 + (1-P_0)I_1]}{I_0}.
\end{equation}
For the single-resonance case, the contrast is $C_s=\frac{1}{2}\frac{I_0-I_1}{I_0}$. In the dual-resonance case, the contrast is $C_d=\frac{2}{3}\frac{I_0-I_1}{I_0}$. The ratio is therefore $C_d/C_s=4/3$. Since the photoelectron-shot-noise-limited sensitivity is proportional to $1/C$ [Eq.~\eqref{eq:psn}], this corresponds to a $4/3$ reduction in the magnetic noise floor. 

To derive the factor of $4/3$ we assumed that the microwave excitation rate was larger than the optical excitation rate. In experiments, we use $20~{\rm mW}$ of microwave power. This corresponds to a microwave Rabi frequency of ${\sim}0.7~{\rm MHz}$ (\ref{sec:SIrabi}) or a spin flip rate of ${\sim}1.4\times10^6~{\rm s^{-1}}$. The optical intensity used in our experiments was $I_{\rm opt}\approx0.2~{\rm W}/(40~{\rm \upmu m})^2=12.5~{\rm kW/cm^2}$ (\ref{sec:SIexpopt}). The NV absorption cross section at $532~{\rm nm}$ is $\sigma_{nv}\approx3\times10^{-17}~{\rm cm^2}$~\cite{ACO2009}, so this corresponds to an optical excitation rate of $I_{\rm opt}\sigma_{\rm nv}/E_{ph}\approx10^6~{\rm s^{-1}}$. Thus, in our experiments, the microwave excitation rate is comparable to, or slightly larger than, the optical excitation rate. The improvement in dual-resonance sensitivity was thus not exactly $4/3$, but it was close (${\sim}1.3)$. Another assumption that we implicitly made is that the FDMR linewidth is the same under single-resonance and dual-resonance excitation. This assumption is reasonably accurate in our experiments, see Fig.~\ref{fig:lock}(c) of the main text.

\begin{center}
\section{\label{sec:SItest}} 
\setlength{\parskip}{-0.8em}{
\textbf {Magnetometer frequency response}}
\end{center}
 
Figure~\ref{fig:results}(c) of the main text shows the amplitude of test fields, recorded by dual-resonance diamond magnetometry, as a function of their frequency. A moderate decay ($\sim40\%$) of the signal amplitude was observed over the $1\mbox{--}1000~{\rm Hz}$ range. In order to determine the causes of this signal decay, we performed a series of frequency-response measurements under different conditions. 

Figure~\ref{fig:test} shows the results of these experiments. In all cases, we use $f_{\rm mod}=15~{\rm kHz}$ and the lock-in uses a 12 dB/octave low-pass filter with a time constant $\tau_{li}=100~{\rm \upmu s}$. We first isolated the lock-in amplifier's frequency response by applying a sinusoidal voltage, oscillating at $f_{\rm mod}=15~{\rm kHz}$, with an amplitude modulation of constant depth and variable modulation frequency. The resulting lock-in response is well described by a second-order Bessel filter with a cutoff frequency of $1/(2\pi\tau_{li})$. While this filter is largely responsible for the magnetometer decay at frequencies ${\gtrsim}1~{\rm kHz}$, it can only account for a small fraction of the decay observed over the $1\mbox{--}1000~{\rm Hz}$ range. 

\begin{figure}[ht]
    \centering
\includegraphics[width=\columnwidth]{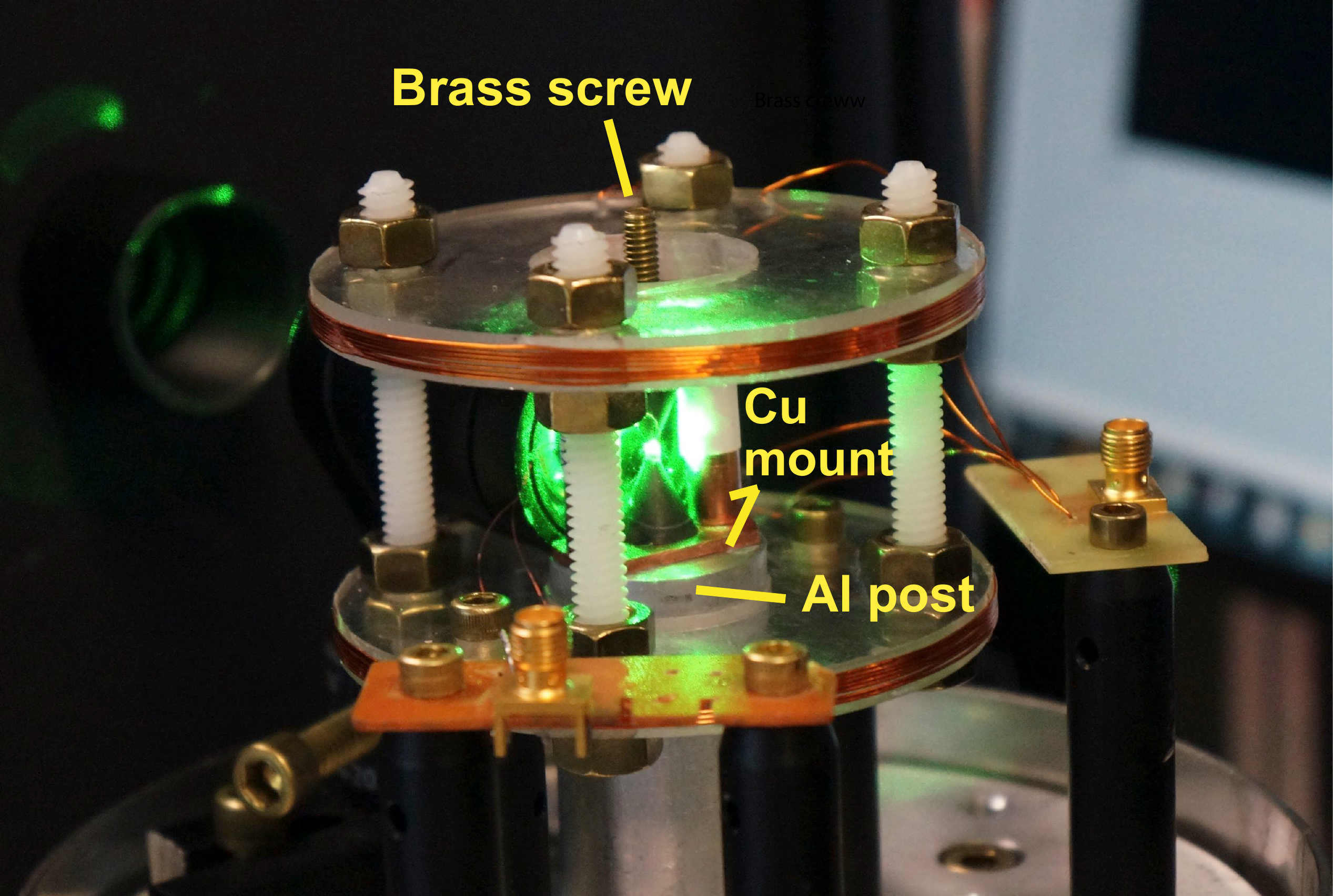}\hfill
       \caption{\label{fig:metals}
\textbf{Photo of apparatus with shield removed.} Metal mounting components that were removed to generate the data in Fig.~\ref{fig:test} are labeled. The brass screw was used for mounting to a translation stage during initial alignment (\ref{sec:SIexpcones}). Other unlabeled metal parts, such as brass nuts, were not found to contribute to the frequency-dependent magnetic field attenuation.}
\end{figure}

Next, we removed the ferrite cones from the assembly and performed dual-resonance magnetometry. The observed frequency response is similar to that observed with the ferrite cones in place. The decay is slightly less pronounced, but evidently the ferrite cones do not account for the observed decay.

Finally, we removed the metal mounting hardware used in the apparatus that were located within the Helmholtz coils, Fig.~\ref{fig:metals}. We again performed dual-resonance diamond magnetometry without the ferrite cones in place. In this case, we observe a frequency response which is nearly identical to the lock-in amplifier's frequency response. 

We therefore conclude that metal components within the Helmholtz coils are responsible for most of the decay in the $1\mbox{--}1000~{\rm Hz}$ range observed in Fig.~\ref{fig:results}(c). The lock-in amplifier's low-pass filter contributes as well, but to a lesser degree. The ferrite cones may also contribute a small amount to the observed decay, but future work would be needed to isolate their response independently.

\begin{figure}[t]
    \centering
\includegraphics[width=\columnwidth]{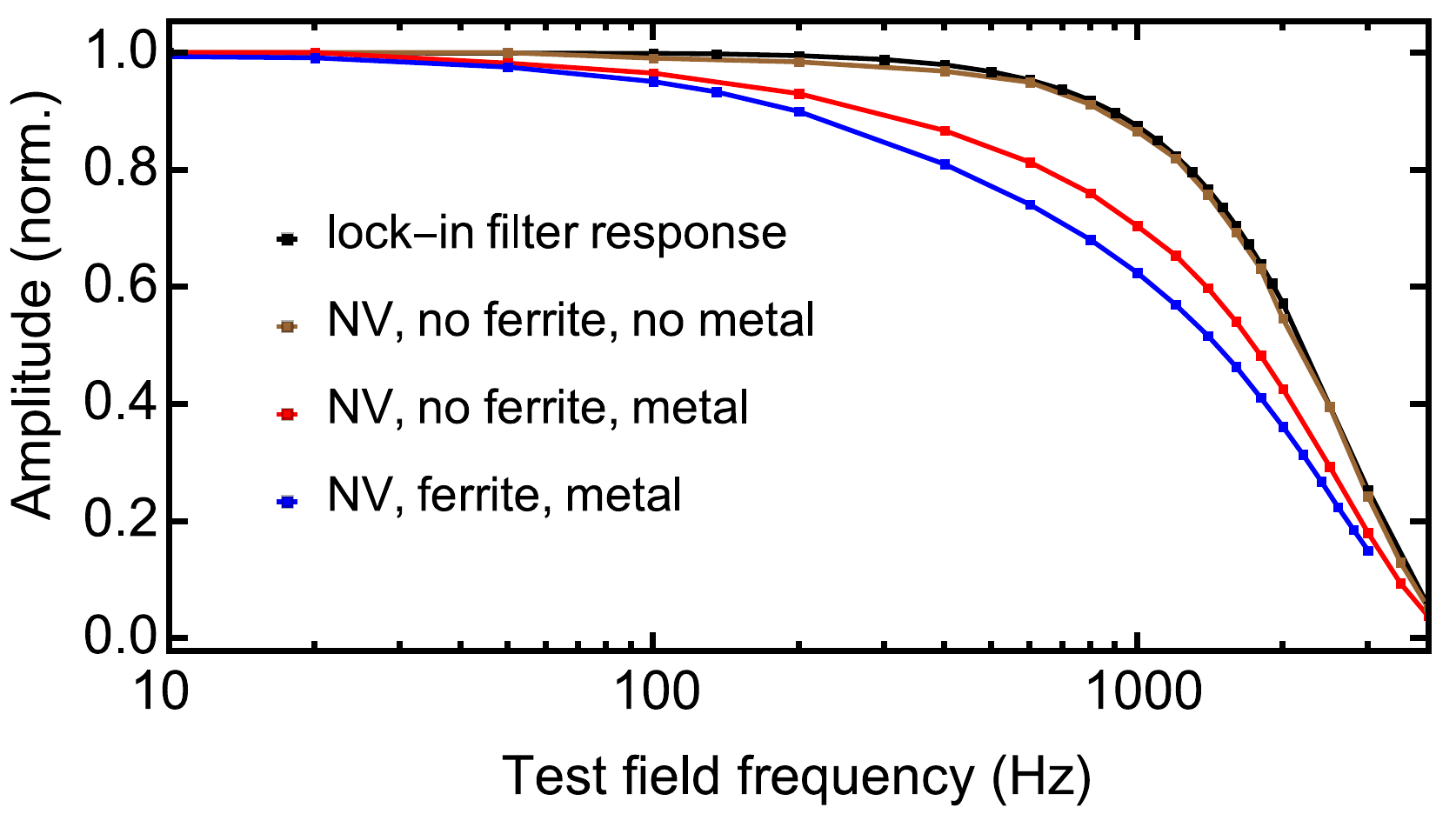}\hfill
       \caption{\label{fig:test}
\textbf{Frequency response of different magnetometer configurations.} The blue trace is the normalized magnetometer frequency response, reproduced from Fig.~\ref{fig:results}(c). The red trace is the same NV magnetometer setup except without the ferrite cones. The brown trace is the NV magnetometer without ferrite cones and with metal components (Fig.~\ref{fig:metals}) removed from the interior of the Helmholtz coils. The black trace is the lock-in filter response as measured by amplitude-modulated voltage inputs. }
\end{figure}

The frequency dependence of our magnetometer leaves an ambiguity as to how best to normalize the magnetic noise spectra in Fig.~\ref{fig:results}(b). As seen in Fig.~\ref{fig:results}(c), when we apply a test current which is expected to produce an amplitude of $580~{\rm pTrms}$, it produces the correct amplitude at 1 Hz, but at $125\mbox{--}135~{\rm Hz}$ it produces an amplitude of ${\sim}540~{\rm pTrms}$. Since $125\mbox{--}135~{\rm Hz}$ is the frequency range of the test fields applied in Fig.~\ref{fig:results}(b), we therefore had to decide whether to normalize the noise spectra so that the test-field peaks appeared at $580~{\rm pT\,s^{1/2}}$ or ${\sim}540~{\rm pT\,s^{1/2}}$. Conservatively, we chose the former. We multiplied each spectrum by $580/540=1.07$, which raised the test field peaks to $580~{\rm pT\,s^{1/2}}$ and also raised the noise floor by $7\%$. If we had instead chosen to normalize the test-field peaks to $540~{\rm pT\,s^{1/2}}$, our noise floor estimates would improve by ${\sim}7\%$ to ${\sim}0.84~{\rm pT\,s^{1/2}}$.

\begin{center}
\section{\label{sec:SIrabi}} 
\setlength{\parskip}{-0.8em}{
\textbf {Ferrite microwave field enhancement}}
\end{center}

A feature of our magnetometer is that it uses a simple, non-resonant coil for microwave excitation and only requires $20~{\rm mW}$ of microwave power. This is partially enabled by an enhancement of the microwave magnetic field provided by the ferrite cones. Figure~\ref{fig:rabi} shows Rabi oscillations of the same diamond-coil configuration with and without ferrite cones. The Rabi frequency with ferrite is a $\gtrsim2$-times larger, indicating an equivalent $\gtrsim2$-fold enhancement in the microwave magnetic field.

\begin{center}
\section{\label{sec:SIcalib}} 
\setlength{\parskip}{-0.8em}{
\textbf {Calibration of Helmholtz coils}}
\end{center}

Our magnetometer signal's accuracy relies on a careful calibration of the conversion between the current applied to the Helmholtz coils and $B_{\rm ext}$. Here, we call this conversion factor $M_{\rm cal}$. Theoretically, we estimated $M_{\rm cal}=165~{\rm \upmu T/A}$ based on the known coil geometry and number of turns. We verified this estimate experimentally by applying currents to the Helmholtz coils and measuring the resulting magnetic field using three different magnetometers.

\begin{figure}[hb]
    \centering
\includegraphics[width=\textwidth]{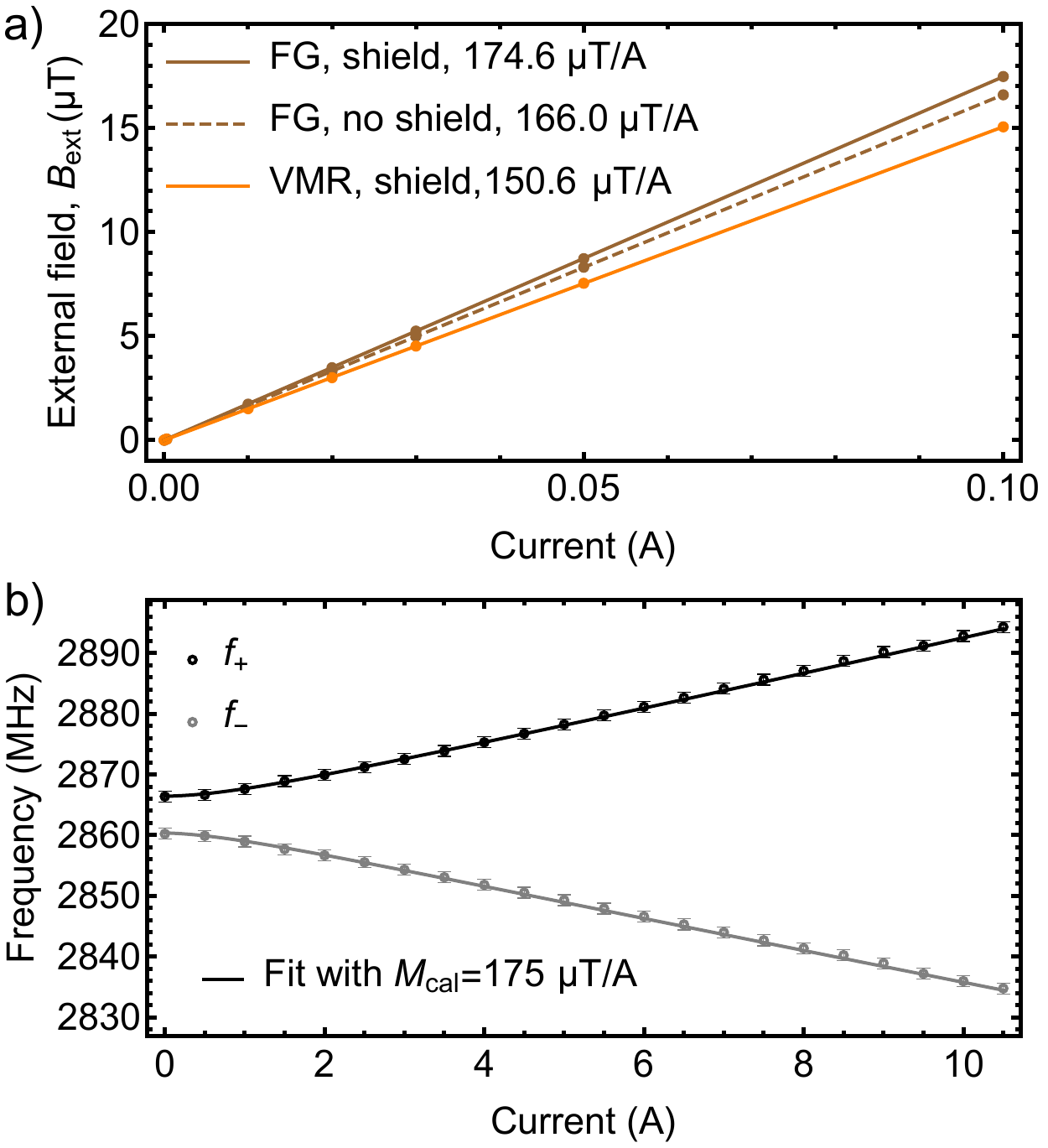}\hfill
       \caption{\label{fig:coils}
\textbf{Helmholtz coils current calibration.} (a) Helmholtz coils current calibration performed with two commercial vector magnetometers. FG: SENSYS FGM3D/100 fluxgate magnetometer;
VMR: Twinleaf VMR magnetometer. (b) NV FDMR frequencies versus current in the Helmholtz coils. Solid lines are a fit using Eq.~\ref{eq:htot}, where $M_{\rm cal}=175~{\rm \upmu T/A}$ is the fit parameter. }
\end{figure}

First, two commercial vector magnetometers (Twinleaf VMR and SENSYS fluxgate, see Fig.~\ref{fig:mags}) were used to calibrate the Helmholtz coils. Each magnetometer was placed in the center of the coils at approximately the same location as the NV-ferrite structure would be. The current in the Helmholtz coils was varied and the axial magnetic field component was recorded. Figure~\ref{fig:coils}(a) shows the resulting calibration curves. The data were fit to linear functions, revealing $M_{\rm cal}$ (listed in the legend). For the fluxgate magnetometer, $M_{\rm cal}$ is approximately the same as the theoretical estimate when the top of the magnetic shields was removed. When the shield remained in place, the calibration factor was $\sim5\%$ larger. The VMR magnetometer reported a lower magnetic field than other methods. In both cases we relied on conversion constants between voltage and magnetic field units as provided by the manufacturers. 

Note that when the current was turned off, we still observed a small residual axial field of $B_{\rm ext}=-0.2~{\rm \upmu T}$ using both magnetometers. This is due to the finite attenuation provided by the shields. When the shields were removed, the axial component of the lab field was approximately $-20~{\rm \upmu T}$. Since the shields provide a ${\sim}100$-fold attenuation, this leads to a small residual field of $-0.2~{\rm \upmu T}$.

Next, we used NV magnetometry, with the ferrite cones removed from the setup (\ref{sec:SItest}), to measure the FDMR frequencies as a function of the current in the coils. Figure~\ref{fig:coils}(b) shows the observed $f_{\pm}$ values alongside a fit according to the NV spin Hamiltonian, Eq.~\ref{eq:htot}, with $M_{\rm cal}=175~{\rm \upmu T/A}$ as a fitting parameter. 

The value of $M_{\rm cal}$ used throughout the main text was the average of all three values reported by the magnetometers with the shields on. It is $M_{\rm cal}=167\pm14~{\rm \upmu T/A}$, where the uncertainty is the standard deviation. If we had removed the VMR magnetometer from the analysis, we would have obtained $M_{\rm cal}\approx175~{\rm \upmu T/A}$. This would decrease the reported sensitivity by ${\sim}5\%$ to $\sim 0.95~{\rm pT\,s^{1/2}}$.    

\begin{figure}[hb]
    \centering
\includegraphics[width=\columnwidth]{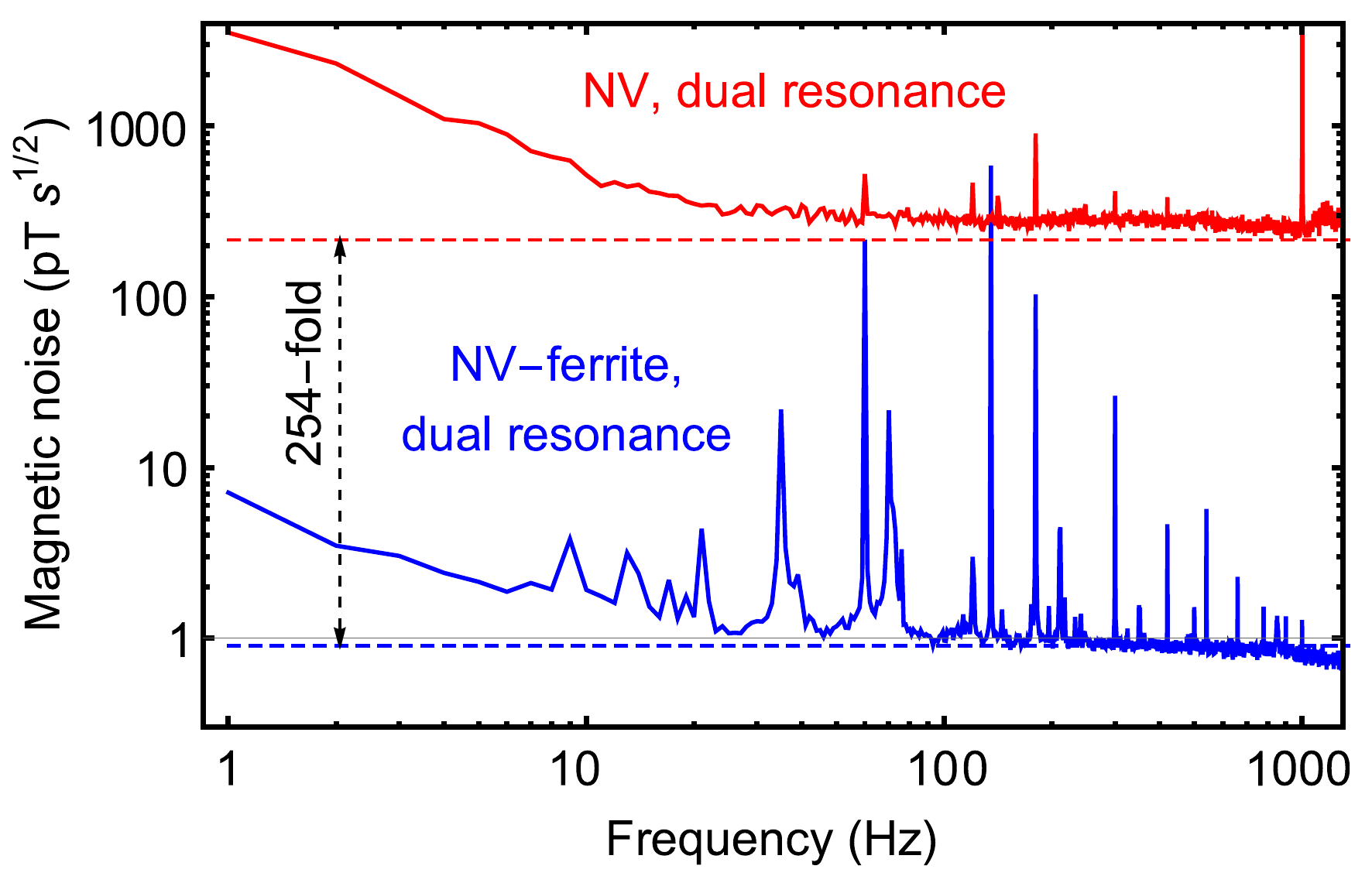}\hfill
       \caption{\label{fig:noferrite}
\textbf{Sensitivity with and without ferrite cones.}  
Magnetic noise spectra for dual-resonance magnetometry with (blue) and without (red) the ferrite cones.}
\end{figure}

\begin{center}
\section{\label{sec:SImagsens}} 
\setlength{\parskip}{-0.8em}{
\textbf {Sensitivity without ferrite}}
\end{center}
We used the same dual-resonance magnetometry technique described in the main text to record the diamond magnetometry signal with the ferrite cones removed. Fig.~\ref{fig:noferrite} shows the resulting magnetic noise spectrum alongside the spectrum with ferrite [reproduced from Fig.~\ref{fig:results}(b)]. The noise floor without ferrite is $\sim300~{\rm pT\,s^{1/2}}$. This is slightly larger than the expected $254$-fold increase, most likely due to a suboptimal choice of microwave power.

\begin{center}
\section{\label{sec:SIhysteresis}} 
\setlength{\parskip}{-0.8em}{
\textbf {Flux concentrator hysteresis}}
\end{center}

The data in Fig.~\ref{fig:setup}(c) of the main text were obtained by sweeping $B_{\rm ext}$ from zero to $+50~{\rm \upmu T}$, then from $+50~{\rm \upmu T}$ to $-50~{\rm \upmu T}$, and finally from $-50~{\rm \upmu T}$ back to zero. To check whether hysteresis results in any remanent fields over the course of these measurements, we separated the $f_{\pm}$ data into three segments: $0\mbox{--}+50~{\rm \upmu T}$, $+50\mbox{--}-50~{\rm \upmu T}$, and $-50\mbox{--}0~{\rm \upmu T}$. We fit the three data sets separately according to Eq.~\ref{eq:htot} with a residual magnetic field offset of $B_{\rm ext}$ as the only fitting parameter. The resulting offset magnetic fields were found to be 8.8 nT, -9.2 nT, and 9.8 nT, respectively. This variation lies within the fit uncertainty, so we take $10~{\rm nT}$ as an upper bound. Note that this corresponds to a remanent field within the gap of $\lesssim2.5~{\rm \upmu T}$.

\begin{center}
\section{\label{sec:SIshot}} 
\setlength{\parskip}{-0.8em}{
\textbf {Photoelectron shot-noise limit}}
\end{center}

The photoelectron-shot-noise-limited sensitivity of our magnetometer is given by:
\begin{equation}
    \label{eq:psnSI}
    \eta_{\rm psn}=\frac{\sqrt{q\,I_{\rm dc}}}{dI_{\rm ac}/dB_{\rm ext}},
\end{equation}
where $I_{\rm dc}=2.3~{\rm mA}$ is the sum of the average photocurrent in both channels of the balanced photodetector, $dI_{\rm ac}/dB_{\rm ext}=33~{\rm Arms/T}$ is the lock-in slope expressed in terms of the AC photocurrent rms amplitude [Fig.~\ref{fig:lock}(c)], and $q=1.6\times10^{-19}~{\rm C}$ is the electron charge. Thus Eq.~\eqref{eq:psnSI} evaluates to $\eta_{\rm psn}=0.58~{\rm pT\,s^{1/2}}$. This noise can be thought of as the standard deviation of the time-domain magnetometer data obtained in 1 second intervals. In the frequency domain it corresponds to the standard deviation of the real part of the Fourier Transform expressed in ${\rm pT\,s^{1/2}}$. In our experiments, we report the absolute value of the Fourier Transform. In order to represent $\eta_{\rm psn}$ in this way, it must be multiplied by 1.25 to reveal a magnetic noise floor of $\eta_{\rm psn}=0.72~{\rm pT\,s^{1/2}}$. This conversion was checked by simulating Poissonian noise and observing the noise floor in the absolute value of the Fourier Transform.

A similar value $\eta_{\rm psn}\approx0.75~{\rm pT\,s^{1/2}}$ was obtained by inserting experimental values from FDMR spectra into Eq.~\ref{eq:psn} in the main text. In this case, we used $\xi=0.01$, $P_{\rm opt}=200~{\rm mW}$, $\Gamma=9~{\rm MHz}$, and $C=0.04$. The effect of flux concentrators is incorporated by multiplying $\gamma_{nv}$ by $\epsilon$. Note that the expression in Eq.~\ref{eq:psn} refers to the sensitivity to the magnetic field component along the NV axis. Since we use this measurement to infer the total field amplitude (which is directed at $55\degree$ with respect to the NV axes), the right hand side of Eq.~\ref{eq:psn} must be multiplied by $1/\cos{55\degree}=\sqrt{3}$. 

In Sec.~\ref{sec:Introduction} of the main text, we claimed that the lowest value of $\Gamma/C$~\cite{BAR2016} was $1~{\rm MHz}/0.04$. To be accurate, the reported contrast in this paper was $0.05$ and the linewidth was $1~{\rm MHz}$. However this experiment measured the projection of the field onto NV axes that were aligned at $35\degree$ with respect to the field (the field was aligned normal to a [110]-cut diamond face). Incorporating the projection factor ($\cos{35\degree}=0.82$) in Eq.~\ref{eq:psn} has the same effect as scaling down the ratio $\Gamma/C$ by the same proportion. We thus reported the ratio as $\Gamma/C\approx1~{\rm MHz}/0.04$.

Finally, we would like to clarify some issues with the optimistic estimation of $\eta_{\rm psn}$ made in Sec.~\ref{sec:Introduction} of the main text. There, we combined the highest-reported value of $\xi$ with the lowest reported value of $\Gamma/C$. In reality such a combination may be difficult to achieve as there are competing factors. For example, realizing high $\xi$ requires high optical depth. This is challenging to realize when $\Gamma$ is small, because the latter implies a low NV density. In principle this could still be accomplished with a multipass configuration or by using a large diamond. However, as one moves to lower $\Gamma$, the optimal excitation intensity also decreases (since the optical excitation rate should not exceed $\Gamma$). This means that, for a fixed power, the beam area must increase, which further constrains the geometry and favors larger diamond dimensions. A lower excitation intensity also results in a smaller magnetometer bandwidth, since the optical repolarization rate is lower. Finally, realizing a high value of $\xi$ requires getting waveguides and/or lenses very close to the diamond. Realizing such a high optical access may interfere with other magnetometer components (concentrators, microwave loop, heat sinks, etc.). Most of these technical challenges are not insurmountable, but they need to be addressed. Our flux concentrator solution offers a complementary path that may alleviate some of these engineering constraints.

\begin{table*}[t]
\caption{\label{tab:noises} \textbf{Thermal magnetic noise for different cone materials.} Magnetic noise arising from Hysteresis and Johnson noise were numerically calculated by the method described in Refs.~\cite{GRI2009,LEE2008} using finite-element methods. The values of $\mu'$, $\mu''$, and $\sigma$ are taken from references: low-carbon steel 1018~\cite{BOW2006}, MnZn ferrite MN80~\cite{KOR2007}, MnZn ferrite MN60~\cite{GRI2009}, and mu-metal~\cite{GRI2009}. Note that $\mu'$ and $\mu''$ are in general frequency dependent. Here we take the values for the lowest reported frequency and assume that the response is relatively flat below 1 kHz. The enhancement $\epsilon$ is determined from magnetostatic simulations as in Fig.~\ref{fig:sim} of the main text. The effective external magnetic noises $\delta B_{\rm ext}=\delta B_{\rm gap}/\epsilon$ are defined by Eqs.~\ref{eq:losses}-\ref{eq:noise}. $\delta B_{\rm ext}$ is reported at $1~{\rm Hz}$. It scales with frequency as $f^{-1/2}$.}
\begin{ruledtabular}
\begin{tabular}{lrrcccr}
  \bf{Material}  & $\pmb{\mu'/\mu_0}$  & $\pmb{ \mu''/\mu_0}$  & $\pmb{\sigma}$ (S/m) & \bf{Enhancement, $\epsilon$} & $\pmb{\delta B_{hyst}}$ {\bf (1 Hz)}, pT s$^{1/2}$ & $\pmb{\delta B_{eddy}}$, pT s$^{1/2}$  \\
  \hline
  Steel 1018 & 250 & 5  & 5.18$\times 10^6$ & 223 & 6.8 & 0.4 \\
  MnZn MN80 & 2030 & 6.1   & 0.2  & 251 & 0.85 & 0.00007 \\
  MnZn MN60 & 6500 & 26   & 0.2  & 254  & 0.54 & 0.00007 \\  
  mu-metal & 30000 & 1200   & 1.6$\times 10^6$  & 255 & 0.8 & 0.2 \\
\end{tabular}
\end{ruledtabular}
\end{table*}

\begin{center}
\section{\label{sec:SInoise}} 
\setlength{\parskip}{-0.8em}{
\textbf {Ferrite thermal magnetic noise}}
\end{center}

Thermal magnetic noise originating from dissipative materials can be estimated using fluctuation-dissipation methods~\cite{GRI2009,LEE2008}. The noise is inferred by calculating the power loss ($P$) incurred in the material due to a hypothetical oscillating magnetic field (angular frequency: $\omega$) produced by a small current loop (area: $A$, current: $I$) situated at the location of the magnetometer. The magnetic noise detected by the sensor is then given by:
\begin{equation}
\label{eq:noise}\delta B_{\rm gap}=\frac{\sqrt{8 kT P}}{A I \omega} ,  
\end{equation}
where $k$ is the Boltzmann constant. The power loss has separate contributions due to thermal eddy currents and magnetic domain fluctuations: 
\begin{equation}
\label{eq:losses}P_{\rm eddy}=\int_V \frac{1}{2} \sigma E^2 d V, \;\; P_{\rm hyst}=\int_V \frac{1}{2} \omega \mu'' H^2 d V. 
\end{equation}
Here $\sigma$ is the electrical conductivity, $\mu''$ is the imaginary part of the permeability ($\mu=\mu'-i\mu''$), $E$ and $H$ are the amplitudes of the induced electric and magnetic fields, and the integration is carried out over the volume $V$ of the dissipative material. In the small excitation limit, $E$ and $H$ scale linearly with the driving dipole moment ($A I$), so the magnetic noise in Eq.~\eqref{eq:noise} is independent of the size and driving current in the loop.

We numerically calculated magnetic noise contributions due to $P_{\rm eddy}$ and $P_{\rm hyst}$ for our flux concentrator geometry [Figs.~\ref{fig:sim}(a-b)]. We used MN60 material parameters~\cite{GRI2009} ($\sigma=0.2~{\rm \Omega^{-1}m^{-1}}$, $\mu'=6500\,\mu_0$, $\mu''=26\,\mu_0$, where $\mu_0$ is the vacuum permeability) and a cone gap of $\delta=47~{\rm \upmu m}$, which resulted in the experimental enhancement factor $\epsilon=254$. We find that thermal eddy currents produce white magnetic noise at the level of $\delta B_{\rm gap}\approx0.02~{\rm pT\,s^{1/2}}$. Since we are interested in our sensitivity in relation to the external field~\cite{GRI2009}, noise produced locally by the ferrite cones translates to an equivalent external field noise of $\delta B_{\rm ext}=\delta B_{\rm gap}/\epsilon\approx7\times10^{-5}~{\rm pT\,s^{1/2}}$. This negligibly-low noise level is a consequence of our choice of low-conductivity ferrite materials. On the other hand, thermal magnetization noise results in a larger, frequency-dependent magnetic noise. At 1 Hz, the effective noise is $0.5~{\rm pT\,s^{1/2}}$, and it scales with frequency as $f^{-1/2}$. The thermal magnetization noise is annotated in Fig.~\ref{fig:results}(b). It is not a limiting factor in our present experiments, but it may have implications for future optimization efforts. If a material with a lower relative loss factor ($\mu''/\mu'^2$) could be identified, it would result in lower thermal magnetization noise (\ref{sec:SInoiseM}).

\begin{center}
\section{\label{sec:SInoiseM}} 
\setlength{\parskip}{-0.8em}{
\textbf {Thermal magnetic noise for various materials}}
\end{center}

We also used Eqs.~\eqref{eq:noise} and \eqref{eq:losses} to estimate the magnetic noise produced by cones of the same dimensions as in Fig.~\ref{fig:sim}(a), but made from different magnetic materials. Specifically, we considered low-carbon steel 1018~\cite{BOW2006}, MnZn ferrite MN80~\cite{KOR2007}, and mu-metal~\cite{GRI2009}. The results of these estimates are listed in Tab.~\ref{tab:noises} along with the material parameters used for the analysis. In all cases, the hysteresis noise is dominant for frequencies $\lesssim10~{\rm Hz}$.

\begin{figure}[hb]
    \centering
\includegraphics[width=\columnwidth]{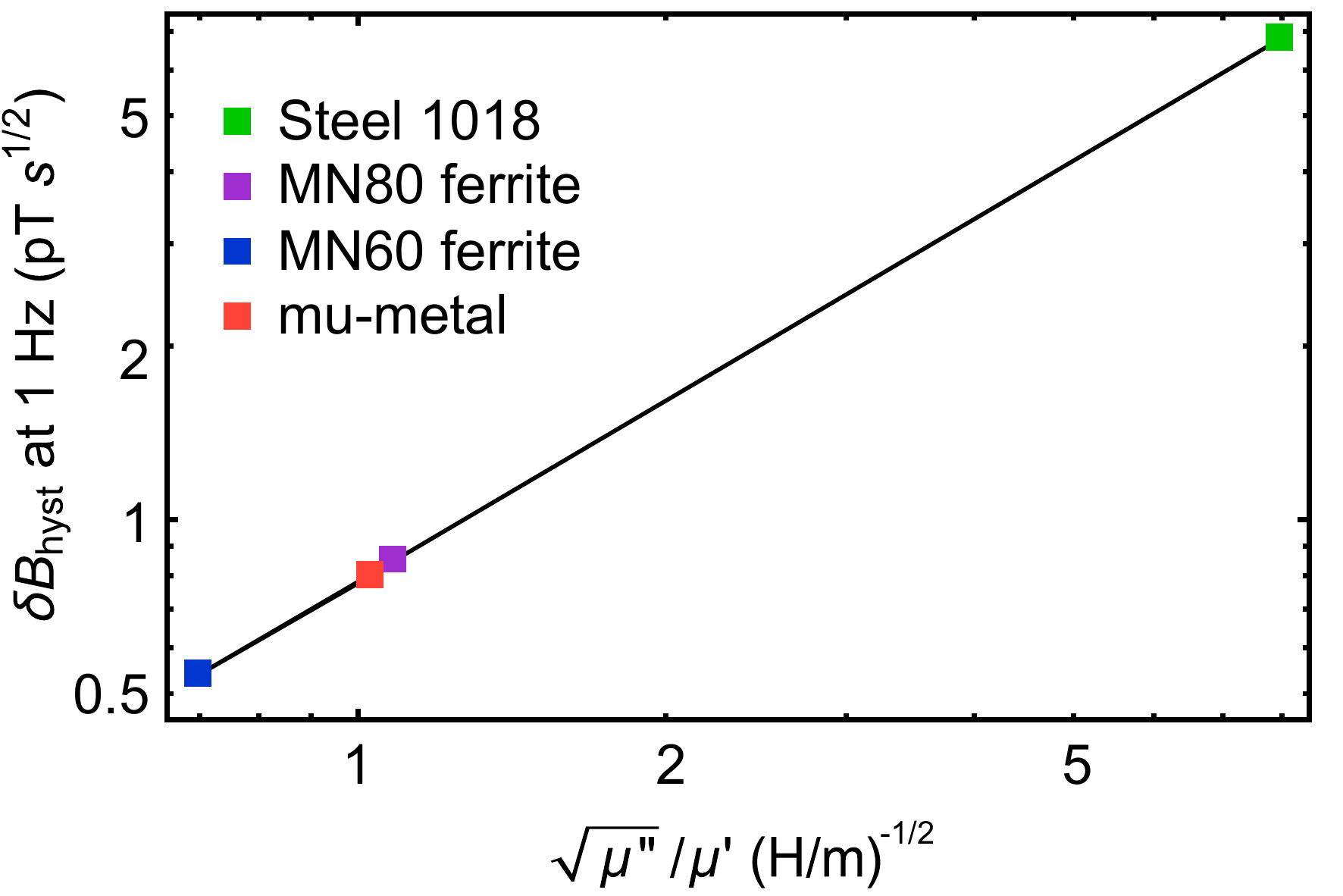}\hfill
       \caption{\label{fig:losses}
\textbf{Magnetization noise vs. relative loss factor.} Calculated hysteresis magnetic noise (at 1 Hz) as a function of the square root of the relative loss factor, $\sqrt{\mu''}/\mu'$ in four magnetic materials. }
\end{figure}

To minimize hysteresis noise, one must limit the relative loss factor ($\mu''/\mu'^2$). We found that the hysteresis noise scales proportional to $\sqrt{\mu''/\mu'^2}$, Fig.~\ref{fig:losses}. Another design consideration is the geometry of the flux concentrators, but such an optimization is beyond the scope of this work. If the Johnson noise matters, as in the conductive mu-metal, it could be further decreased by passivating the skin effect  with a lamination. 

Finally, we estimated the magnetic noise produced by the mu-metal magnetic shield used in our experiments. Here, we used an analytical expression for a finite closed cylinder~\cite{LEE2008} and inserted mu-metal parameters from Tab.~\ref{tab:noises} along with the shield dimensions (height: $150~{\rm mm}$, diameter: $150~{\rm mm}$, thickness: $1.5~{\rm mm}$). The calculated Johnson noise for our shield is $\delta B_{eddy}=0.02~{\rm pT\,s^{1/2}}$ and the hysteresis noise at $1~{\rm Hz}$ is $\delta B_{hyst}=0.007~{\rm pT\,s^{1/2}}$. These are much lower than the observed noise floors and can safely be neglected. Note that the noise from the shields is enhanced by the flux concentrators. This effect was incorporated in the calculations, but we still arrived at negligibly-low values.

\begin{center}
\section{\label{sec:SIgap}} 
\setlength{\parskip}{-0.8em}{
\textbf {Sensitivity to variation of the gap length}}
\end{center}

An important systematic effect in our device could arise from temporal variations in the cone gap length. According to the data in Fig.~\ref{fig:sim}(d), a small change in gap length in the vicinity of $\delta\approx43~{\rm \mu m}$ produces a change in the enhancement factor given by $d\epsilon/d\delta\approx6/\upmu {\rm m}$. This variation extrapolates to a variation in the magnetometer reading given by:
\begin{equation}
\label{eq:gap}\frac{dB_{\rm ext}}{d\delta}=\frac{d\epsilon}{d\delta}\frac{B_{\rm ext}}{\epsilon}. 
\end{equation}
For $\epsilon\approx254$ and $B_{\rm ext}=2~{\rm \upmu T}$, Eq.~\eqref{eq:gap} predicts that a change in $\delta$ of just $20~{\rm pm}$ produces an error in estimation of $B_{\rm ext}$ of $1~{\rm pT}$. 

Fortunately the gap length remains relatively stable in our construction such that this effect may only be a problem at low frequencies. If the material in the gap expands and contracts due to changes in temperature, this produces a thermal dependence of the magnetometer output given by:
\begin{equation}
\label{eq:temp}\frac{dB_{\rm ext}}{dT}=\frac{dB_{\rm ext}}{d\delta}\frac{d\delta}{dT}.
\end{equation} 

If the material in the gap has a thermal expansion coefficient $\alpha$, then the temperature dependence of the gap length is $d\delta/dT=\alpha \delta$. For diamond, $\alpha\approx0.7\times10^{-6}{\rm K^{-1}}$. Using this value, and inserting Eq.~\eqref{eq:gap} into Eq.~\eqref{eq:temp}, we find a magnetometer temperature dependence of $dB_{\rm ext}/dT\approx1.5~{\rm pT/K}$. This temperature dependence is more than 6 orders of magnitude smaller than the thermal dependence in single-resonance magnetometry (Sec.~\ref{sec:exp}). Nevertheless, to reach this limit, care must be taken to mechanically stabilize the gap using an approach which does not significantly increase $d\delta/dT$. For example, using mechanical clamping and/or very thin adhesive layers would be beneficial.

\bibliographystyle{apsrev4-1}
\end{document}